\documentclass[aps,prd,preprint,superscriptaddress,showpacs]{revtex4}
\usepackage{graphicx}
\newcommand{\bm}[1]{\mbox{\boldmath $#1$}}
\newcommand{\be}{\begin{equation}}
\newcommand{\ee}{\end{equation}}
\newcommand{\bea}{\begin{eqnarray}}
\newcommand{\eea}{\end{eqnarray}}

\newcommand{\bfk}{\mbox{\boldmath $k$}}
\newcommand{\bfq}{\mbox{\boldmath $q$}}
\newcommand{\bfP}{\mbox{\boldmath $P$}}

\newcommand{\bfx}{\mbox{\boldmath $x$}}
\newcommand{\bfy}{\mbox{\boldmath $y$}}
\newcommand{\bfz}{\mbox{\boldmath $z$}}
\newcommand{\bfu}{\mbox{\boldmath $u$}}
\newcommand{\bfv}{\mbox{\boldmath $v$}}
\newcommand{\bfX}{\mbox{\boldmath $X$}}
\newcommand{\bfY}{\mbox{\boldmath $Y$}}
\newcommand{\bfZ}{\mbox{\boldmath $Z$}}

\newcommand{\pup}{p^\uparrow}

\newcommand{\qup}{q^\uparrow}
\newcommand{\aup}{a^\uparrow}
\newcommand{\cupar}{c^\uparrow}

\def\slash{\rlap{/}}
\newcommand{\bfp}{\mbox{\boldmath $p$}}

\newcommand{\Aup}{A^\uparrow}

\newcommand{\la}{\lambda}
\newcommand{\hf}{\hat f}
\newcommand{\ua}{\uparrow}
\newcommand{\da}{\downarrow}
\def\lsim{\mathrel{\rlap{\lower4pt\hbox{\hskip1pt$\sim$}}\raise1pt\hbox{$<$}}}
\def\gsim{\mathrel{\rlap{\lower4pt\hbox{\hskip1pt$\sim$}}\raise1pt\hbox{$>$}}}
\def\nostrocostruttino#1\over#2{\mathrel{\mathop{\kern 0pt \rlap
{\hbox{$#1$}}} \hbox{\kern-.135em $#2$}}}
\def\sumint{\nostrocostruttino \sum \over {\displaystyle\int}}
\def\lsim{\mathrel{\rlap{\lower4pt\hbox{\hskip1pt$\sim$}}\raise1pt\hbox{$<$}}}
\def\gsim{\mathrel{\rlap{\lower4pt\hbox{\hskip1pt$\sim$}}\raise1pt\hbox{$>$}}}
\def\nostrocostruttino#1\over#2{\mathrel{\mathop{\kern 0pt \rlap
{\hbox{$#1$}}} \hbox{\kern-.125em $#2$}}}
\def\sumint{\nostrocostruttino \sum \over {\displaystyle\int}}

\newcommand{\SA}{S_{A}}
\newcommand{\SB}{S_{B}}

\newcommand{\fgAS}{\hat{f}_{g/A,\SA}}

\newcommand{\vphi}{\varphi}

\newcommand{\IoneG}{\mathcal{T}_{1}^{g}}
\newcommand{\ItwoG}{\mathcal{T}_{2}^{g}}
\newcommand{\Ione}{\mathcal{T}_{1}}
\newcommand{\Itwo}{\mathcal{T}_{2}}

\newcommand{\NP}[1]{{\it Nucl.\ Phys.}\ {\bf #1}}

\newcommand{\PL}[1]{{\it Phys.\ Lett.}\ {\bf #1}}
\newcommand{\PR}[1]{{\it Phys.\ Rev.}\ {\bf #1}}
\newcommand{\PRL}[1]{{\it Phys.\ Rev.\ Lett.}\ {\bf #1}}

\newcommand{\JHEP}[1]{{\it J.\ High\ Energy\ Phys.}\ {\bf #1}}

\def\kt{k_\perp}
\def\bkt{\bfk_\perp}

\begin{document}
%%%%%%%%%%%%%%%%%%%%%%%%%%%%%%%%%%%%%%%%%%%%%%%%%%%%%%%%%%%%%%%%%%%%%%%%%%%%%%

\title{General partonic structure for hadronic spin asymmetries}

\author{M.~Anselmino}
\affiliation{Dipartimento di Fisica Teorica, Universit\`a di Torino and \\
          INFN, Sezione di Torino, Via P. Giuria 1, I-10125 Torino, Italy}
\author{M.~Boglione}
\affiliation{Dipartimento di Fisica Teorica, Universit\`a di Torino and \\
          INFN, Sezione di Torino, Via P. Giuria 1, I-10125 Torino, Italy}
\author{U.~D'Alesio}
\affiliation{INFN, Sezione di Cagliari and Dipartimento di Fisica,
Universit\`a di Cagliari,\\
C.P. 170, I-09042 Monserrato (CA), Italy}
\author{E.~Leader}
\affiliation{Imperial College London, Prince Consort Road, London SW7 2BW,
U.K.}
\author{S.~Melis}
\affiliation{INFN, Sezione di Cagliari and Dipartimento di Fisica,
Universit\`a di Cagliari,\\
C.P. 170, I-09042 Monserrato (CA), Italy}
\author{F.~Murgia}
\affiliation{INFN, Sezione di Cagliari and Dipartimento di Fisica,
Universit\`a di Cagliari,\\
C.P. 170, I-09042 Monserrato (CA), Italy}

\date{\today}

\begin{abstract}
\noindent
The high energy and large $p_T$ inclusive polarized process, 
$(A, S_A) + (B, S_B) \to C + X$, is considered under the assumption of 
a generalized QCD factorization scheme. For the first time all transverse
motions, of partons in hadrons and of hadrons in fragmenting partons, are
explicitly taken into account; the elementary interactions are
computed at leading order  
with noncollinear exact kinematics, which introduces many phases in the
expressions of their helicity amplitudes. Several new spin and $\bfk_\perp$
dependent soft functions appear and contribute to the cross sections and to
spin asymmetries; we put emphasis on their partonic interpretation, in terms
of quark and gluon polarizations inside polarized hadrons. Connections with
other notations and further information are given in some Appendices. The
formal expressions for single and double spin asymmetries are derived. 
The transverse single spin asymmetry $A_N$, for $\pup \, p \to \pi \, X$ 
processes is considered in more detail, and all contributions are
evaluated numerically by saturating unknown functions with their upper
positivity bounds. It is shown that the integration of the phases arising from 
the noncollinear kinematics strongly suppresses most contributions to the 
single spin asymmetry, leaving at work predominantly the Sivers effect and, 
to a lesser extent, the Collins mechanism.
\end{abstract}

\pacs{13.88.+e, 13.60.-r, 13.15.+g, 13.85.Ni}

\maketitle

\section{\label{Intro} Introduction and formalism}
There is, at present, no completely rigorous theory of single spin asymmetries 
in hadron--hadron collisions and inclusive particle production. 
Rigorous results about how different physical
processes, including hadronic ones, are related to each other via 
factorization, only exist for the restricted case of collinear kinematics. 
But it is precisely in this kinematic situation that one cannot generate 
single spin asymmetries at leading twist. 
Thus, the introduction of intrinsic $k_\perp$ is 
crucial for a model of single spin asymmetries and we are therefore forced 
to rely on an intuitively reasonable calculational approach, within QCD, 
assuming a simple factorization scheme. This effectively neglects the role 
of the soft factors related to the Wilson lines which occur in the rigorous 
definition of $k_\perp$ dependent parton densities and fragmentation 
functions.  

In recent papers \cite{fu, noi} we have discussed such a formalism to 
compute cross sections for polarized and
unpolarized inclusive processes, $A \, B \to C\,X$, fully taking into account
parton intrinsic motion in distribution and fragmentation functions, as well
as in the elementary dynamics. In particular, in Ref.~\cite{noi} the emphasis
was on the importance of the many phases appearing in the computation of
helicity amplitudes in noncollinear configurations, and their role in
suppressing the contribution of the Collins mechanism \cite{col} to transverse 
single spin asymmetries. Many other contributions to polarized and unpolarized
cross sections, and to single and double spin asymmetries, were not discussed, 
referring to a later paper for the full treatment of the most complete case.

We consider here such a general case. Let us start from Eq.~(8) of
Ref.~\cite{noi}:
\bea
\frac{E_C \, d\sigma^{(A,S_A) + (B,S_B) \to C + X}}
{d^{3} \bfp_C} = \!\!\!\!\! \sum_{a,b,c,d, \{\la\}}
&& \!\!\! \int \frac{dx_a \,
dx_b \, dz}{16 \pi^2 x_a x_b z^2  s} \;
d^2 \bfk_{\perp a} \, d^2 \bfk_{\perp b}\, d^3 \bfk_{\perp C}\,
\delta(\bm{k}_{\perp C} \cdot \hat{\bm{p}}_c) \, J(\bm{k}_{\perp C})
\nonumber \\
&\times& \rho_{\la^{\,}_a,
\la^{\prime}_a}^{a/A,S_A} \, \hat f_{a/A,S_A}(x_a,\bfk_{\perp a})
\> \rho_{\la^{\,}_b, \la^{\prime}_b}^{b/B,S_B} \,
\hat f_{b/B,S_B}(x_b,\bfk_{\perp b}) \label{gen1} \\
&\times& \hat M_{\la^{\,}_c, \la^{\,}_d; \la^{\,}_a, \la^{\,}_b} \,
\hat M^*_{\la^{\prime}_c, \la^{\,}_d; \la^{\prime}_a,
\la^{\prime}_b} \> \delta(\hat s + \hat t + \hat u) \> \hat
D^{\la^{\,}_C,\la^{\,}_C}_{\la^{\,}_c,\la^{\prime}_c}(z,\bfk_{\perp
C}) \>, \nonumber
\eea
which gives the cross section for the polarized hadronic process
$(A,S_A) + (B,S_B) \to C + X$ as a (factorized) convolution of all possible
hard elementary QCD processes, $ab \to cd$, with soft partonic polarized
distribution and fragmentation functions. In Eq.~(\ref{gen1}) $\hat s, \hat t$
and $\hat u$ are the Mandelstam variables for the partonic reactions and
the detailed connection between the hadronic and the partonic kinematical 
variables is given in full in Appendix A. 

Let us clarify the physical meaning of Eq.~(\ref{gen1}) -- our starting 
point -- by making detailed comments on its notation and contents.

\begin{itemize}
\item
$A$ and $B$ are initial spin 1/2 hadrons (typically, two protons), in pure
spin states denoted by $S_A$ and $S_B$ respectively, with corresponding
polarization vectors $\bfP^A$ and $\bfP^B$ (notice that $\bfP^{A,B}$ are 
actually pseudovectors). We set $S_{A,B} = 0$ for unpolarized hadrons 
($\bfP^{A,B} =$ {\bf 0}). $E_C$ and $\bfp_C$ are the energy and three-momentum 
of the final detected particle (typically, a pion). Throughout the paper, 
we work in the $AB$ c.m. frame, assuming that hadron $A$ moves along the 
positive $Z_{cm}$-axis and hadron $C$ is produced in the $(XZ)_{cm}$ 
plane, with $(p_C)_{X_{cm}}>0$. We define as transverse polarization for 
hadrons $A$ and $B$ the $Y_{cm}$-direction, often using the notation
\be
\uparrow \; {\rm for} \; P_{Y_{cm}}^{A} = 1 \; {\rm and}  \;
P_{Y_{cm}}^{B} = -1
\quad\quad\quad\quad
\downarrow \; {\rm for} \; P_{Y_{cm}}^{A} = -1 \; {\rm and}  \;
P_{Y_{cm}}^{B} = 1
\>.
\ee

The longitudinal spin states are labelled by their helicities:
$\lambda_{A,B} = \pm 1/2$ (sometimes just written as $\pm$) corresponding
to $P_{Z_{cm}}^A = \pm 1$ and  to $P_{Z_{cm}}^B=\mp 1$ respectively. The 
opposite signs for hadrons $A$ and $B$ originate from the fact that their 
helicity frames, as reached from the overall c.m. frame, have opposite
$Y$ and $Z$ axes \cite{elliot}, see Eq.~(\ref{helB}). 
The general case of hadrons transversely polarized along a generic direction
$\phi_{S_A}$ in the $(XY)_{cm}$ plane is treated in Appendix B.
\item
The notation $\{\lambda\}$ implies a sum over {\it all} helicity indices.
$x_a$, $x_b$ and $z$ are the usual light-cone momentum fractions, of
partons in hadrons ($x_{a,b}$) and hadrons in partons ($z$).
$\bfk_{\perp a} \,(\bfk_{\perp b})$ and $\bfk_{\perp C}$ are respectively
the transverse momenta of parton $a\,(b)$ with respect to hadron $A \,(B)$,
and of hadron $C$ with respect to parton $c$. We consider all partons as 
massless, neglecting heavy quark contributions.
\item
With massless partons, the function $J$ is given by \cite{fu}
\be\label{jacobian}
J(\bm{k}_{\perp C}) =
\frac{\left( E_C + \sqrt{\bm{p}^2_C - \bm{k}^2_{\perp C}} \right)^2}
{4(\bm{p}^2_C - \bm{k}^2_{\perp C})}
\>\cdot
\ee
\item
$\rho_{\la^{\,}_a, \la^{\prime}_a}^{a/A,S_A}$ is the helicity density matrix
of parton $a$ inside the polarized hadron $A$, with spin state $S_A$.
Similarly for parton $b$ inside hadron $B$ with spin $S_B$. Notice that the
helicity density matrix describes the spin orientation of a particle in
{\it its helicity frame}~\cite{elliot}; for a spin 1/2 particle,
Tr$\,(\sigma_i \rho) = P_i$ is the $i$-component of the polarization vector
$\bfP$ in the helicity rest frame of the particle. Obviously, for a massless
parton there is no rest frame and the helicity frame is defined as
the standard frame \cite{elliot} in which its four-momentum is
$p^\mu = (p, 0, 0, p)$ (see also Appendix D).
$\hat f_{a/A,S_A}(x_a,\bfk_{\perp a})$ is the distribution function of the
unpolarized parton $a$ inside the polarized hadron $A$. We shall also denote
by $\hat f^a_{s_i/S_J}$ the number densities of partons $a$, with spin along
the $i$-axis, inside a hadron $A$ with spin along the $J$-axis: $i=x,y,z$
stand for directions in the parton helicity frame, whereas $J=X,Y,Z$ refer to
the hadron helicity rest frame.
\item
The $\hat M_{\la^{\,}_c, \la^{\,}_d;
\la^{\,}_a, \la^{\,}_b}$'s are the helicity amplitudes for the
elementary process $ab \to cd$, normalized so that the unpolarized
cross section, for a collinear collision, is given by
\be
\frac{d\hat\sigma^{ab \to cd}}{d\hat t} = \frac{1}{16\pi\hat s^2}\frac{1}{4}
\sum_{\la^{\,}_a, \la^{\,}_b, \la^{\,}_c, \la^{\,}_d}
|\hat M_{\la^{\,}_c, \la^{\,}_d; \la^{\,}_a, \la^{\,}_b}|^2\,.
\label{norm}
\ee
\item
$\hat D^{\la^{\,}_C,\la^{\prime}_C}_{\la^{\,}_c,\la^{\prime}_c}(z,
\bfk_{\perp C})$ is the product of {\it fragmentation amplitudes} for the
$c \to C + X$ process
\be
\hat D^{\la^{\,}_C,\la^{\prime}_C}_{\la^{\,}_c,\la^{\prime}_c}
= \> \sumint_{X, \la_{X}} {\hat{\cal D}}_{\la^{\,}_{C},\,\la^{}_X;
\la^{\,}_c} \, {\hat{\cal D}}^*_{\la^{\prime}_C,\,\la^{}_{X}; \la^{\prime}_c}
\, ,
\ee
where the $\sumint_{X, \la_{X}}$ stands for a spin sum and phase
space integration over all undetected particles, considered as a
system $X$. The usual unpolarized fragmentation function
$D_{C/c}(z)$, {\it i.e.} the number density
 of hadrons $C$ resulting from the fragmentation of an unpolarized
parton $c$ and carrying a light-cone momentum fraction $z$, is given by
\be
D_{C/c}(z) = \frac{1}{2} \sum_{\la^{\,}_c,\la^{\,}_C} \int d^2\bfk_{\perp C}
\, \hat D^{\la^{\,}_C,\la^{\,}_C}_{\la^{\,}_c,\la^{\,}_c}(z, \bfk_{\perp C})
\,. \label{fr}
\ee
\end{itemize}

Eq.~(\ref{gen1}) is written in a factorized form, separating the soft, long 
distance from the hard, short distance contributions. The hard part 
is computable in perturbative QCD (pQCD), 
while information on the soft one has to be extracted 
from other experiments or modeled. As already mentioned and discussed in
Ref.~\cite{noi}, such a factorization with noncollinear kinematics has never 
been formally proven. Indeed, studies of factorization \cite{piet,metz,ji}, 
comparing semi-inclusive deep inelastic scattering (SIDIS) 
and Drell-Yan reactions have indicated unexpected 
modifications of simple factorization, and the situation for inclusive 
particle production in hadron--hadron collisions is not yet resolved. 
Thus, our approach can only be considered as 
a reasonable phenomenological model. Of course, the perturbative calculation 
of the hard part is only reliable if the hard scale -- in this case the square 
of the transverse momentum of the final hadron, $p_T^2$ -- is large enough;
in our case $p_T^2 \ge 2.25$ (GeV/$c$)$^2$.    
It turns out that the data demand \cite{fu} an average value of $k_\perp^2
\simeq 0.64$ (GeV/$c$)$^2$ for the intrinsic transverse momentum of the parton 
distributions. This is relatively small compared to 2.25 (GeV/$c$)$^2$, but
complications can arise from the tail of the Gaussian distribution, as was
discussed in Ref.~\cite{fu} and will be commented on in Section V. 

The intrinsic motion arises both from parton confinement and from QCD gluon 
emission. In that, our approach, based on perturbative computations performed 
at LO in the strong coupling constant, with noncollinear kinematics, could 
partially and effectively contain some of the effects related to soft gluon 
emissions and the threshold resummation of large logarithmic perturbative 
corrections, recently performed within proper collinear factorization 
\cite{dfv}. A study of weighted single spin asymmetries for double-inclusive 
production in hadron-hadron collisions, based on $\bfk_\perp$ factorization 
using a diagrammatic approach, has appeared very recently \cite{piet}.

In the next Section we discuss in detail the soft contributions to
Eq.~(\ref{gen1}), related to parton distribution and fragmentation functions,
while in Section III we give the explicit analytical expressions of all 
elementary amplitudes, convoluted with the corresponding soft functions.
Some contributions to the unpolarized cross section and the transverse 
Single Spin Asymmetry (SSA) are analytically discussed in Section IV. 
Numerical estimates of the maximal contributions of the different spin 
mechanisms, both to the cross section and the transverse SSA, are presented 
and discussed in Section V. General conclusions and comments are given in 
Section VI. Finally, the full noncollinear partonic kinematics and its 
relation with the overall hadronic variables is discussed, for convenience and 
completeness, in Appendix A; the formal relationships between the hadron and 
the parton polarization are widely studied in Appendix B, and the connection 
with other formalims is explicitely worked out in Appendix C. Useful 
definitions of helicity frames are given in Appendix D. 

\section{\label{dist-fn} Soft physics}

Although Eq.~(\ref{gen1}) has already a clear physical interpretation,
we would like to express the parton density matrix elements in terms of parton
polarizations, so that, when performing the helicity sums, each term has
a direct partonic meaning.

Notice that the parton polarizations are, of course, related to their parent 
hadron polarizations. The way the hadron spin is transferred to the partons 
can be formally described, in general, by bilinear combinations of 
the helicity amplitudes for the process $A \to a + X$ (distribution 
amplitudes) \cite{noi,noi95}. Therefore, one could equally well 
interpret Eq.~(\ref{gen1}) either in terms of parton polarizations or in 
terms of the distribution amplitudes. We follow here the former 
approach, which is somewhat more direct. However, the latter approach offers
a deeper understanding of some of the basic properties of our factorized scheme
({\it e.g.} the parity properties) and allows a direct comparison with other 
formalims used to describe the same spin effects. In Appendix B we give the 
full correspondence between parton polarizations and the distribution 
amplitudes, and in Appendix C we derive the explicit relations between our 
formalism and that of the Amsterdam group \cite{amst}.    

\subsection{\label{quark-sect} Quark polarizations}

The helicity density matrix of quark $a$ can be written in terms of the quark
polarization vector components, $\bfP ^a = (P ^a_x, P ^a_y,P ^a_z) =
(P_T^a \cos\phi_{s_a}, P_T^a \sin\phi_{s_a}, P_L^a)$, as
\be
\rho_{\la^{\,}_a, \la^{\prime}_a}^{a/A,S_A} =
{\left(
\begin{array}{cc}
\rho_{++}^{a} & \rho_{+-}^{a} \\
\rho_{-+}^{a} & \rho_{--}^{a}
\end{array}
\right)}_{\!\!\!\!A,S_A} \!\!\!\!\!\!\!
=
\frac{1}{2}\,{\left(
\begin{array}{cc}
1+P^a_z & P^a_x - i P^a_y \\
 P^a_x + i P^a_y & 1-P^a_z
\end{array}
\right)}_{\!\!\!\!A,S_A} \!\!\!\!\!\!\!
=
\frac{1}{2}\,{\left(
\begin{array}{cc}
1+P^a_L & P^a_T \, e^{-i\phi_{s_a}} \\
P^a_T \, e^{i\phi_{s_a}} & 1-P^a_L
\end{array}
\right)}_{\!\!\!\!A,S_A}\!\!\! ,
\label{rho-a}
\ee
where, as explained above, the $x,y$ and $z$-directions are those of the 
helicity frame of parton $a$. Eq.~(\ref{rho-a}) satisfies the well known 
general properties:
\bea
&& \rho^a_{++} + \rho^a_{--} = 1 \label{rhoa+} \\
&& \rho^a_{++} - \rho^a_{--} = P^a_z = P_L^a  \label{rhoa-} \\
&& 2\,{\rm Re} \rho^a_{-+} = 2\,{\rm Re} \rho^a_{+-} = P^a_x
= P_T^a \, \cos\phi_{s_a} \label{rerhoa} \\
&& 2\,{\rm Im} \rho^a_{-+} = -2\,{\rm Im} \rho^a_{+-} = P^a_y
= P_T^a \, \sin\phi_{s_a} \label{imrhoa} \,.
\eea

When performing the sum over the helicity indices $\la^{\,}_a, \la^{\prime}_a$
and $\la^{\,}_b, \la^{\prime}_b$ in Eq.~(\ref{gen1}), one obtains products of
terms of the form
\be
(P^a_j \, \hat f_{a/A,S_A}) =
\hat f^a_{s_j/S_A} - \hat f^a_{-s_j/S_A}
\equiv \Delta \hat f ^a_{s_j/S_A} \>, \label{defdelta}
\ee
where $j=x,y,z$. Similarly for parton $b$ inside hadron $B$.
We use the notations:
\bea
&& (P^a_j \, \hf _{a/A,S_Y}) = \Delta \hf_{s_j/S_Y}^a =
\hf _{s_j/\ua}^a -  \hf _{-s_j/\ua}^a \equiv
\Delta \hf_{s_j/\ua}^a(x_a, \bfk_{\perp a}) \label{DxY}\\
&& (P^a_j \, \hf _{a/A,S_Z}) = \Delta \hf_{s_j/S_Z}^a =
\hf _{s_j/+}^a -  \hf _{-s_j/+}^a \equiv
\Delta \hf_{s_j/+}^a(x_a, \bfk_{\perp a}) \label{DxZ}\\
&& (\hf _{a/A,S_Y}) = \hf_{a/A}(x_a, k_{\perp a}) + \frac{1}{2}\,
\Delta \hf_{a/S_Y}(x_a, \bfk_{\perp a})\,\label{Dunp}.
\label{main-table}
\eea
These amount to eight independent quantities, which represent the
($\bkt$ unintegrated) distribution functions of partons $a \, (= q, \bar q)$
with polarization $\bfP ^a$ (defined in the partonic helicity frame) inside
hadron $A$ with spin $S_A$ (specified in the hadronic helicity frame). All
of these functions have a simple direct physical meaning: for instance, the
$x$-component of Eq.~(\ref{DxY}) -- $(P^a_x \, \hf _{a/A,S_Y})$ -- represents
the amount of polarization along the $x$-axis (in the partonic helicity frame)
carried by partons $a$ inside a transversely polarized hadron
$(A, S_Y)$; $(P^a_y \, \hf _{a/A,S_Y})$ is related to 
the $\bkt$ dependent transversity
distribution, which, upon integration over $d^2\bkt$, gives the familiar
{\it transversity} function $h^q_1(x)$ or $\Delta _T  q(x)$ (see also
Appendix B). Similarly,
the $z$-component of Eq.~(\ref{DxZ}) -- $(P^a_z \, \hf _{a/A,S_Z})$ -- is the
unintegrated helicity distribution, which, once integrated over the transverse
momentum, gives the usual helicity distribution $\Delta q(x)$ or $g^q_1(x)$.

Notice that two independent distribution functions appear in the definition 
of $\hf_{a/A,S_Y}$, which is the only term in the sum over 
$\la^{\,}_a, \la^{\prime}_a$ which corresponds to parton $a$ being unpolarized:
$\hf_{a/A}(x_a, k_{\perp a})$, the unintegrated number density of unpolarized 
partons $a$ inside the unpolarized proton $A$, and $\Delta \hf_{a/S_Y}$, the 
Sivers function \cite{siv}. The latter permits the number density of 
unpolarized partons $a$ to depend upon the transverse polarization of the 
parent hadron $A$. In general, for a hadron $A$ in a pure spin state $S_A$ and 
corresponding unit polarization vector $\hat{\bfP}^A$, one has:
\bea
\Delta \hf_{a/S_A}\,(x_a, \bfk_{\perp a}) &\equiv& 
\hf _{a/S_A}\,(x_a, \bfk_{\perp a}) - \hf _{a/-S_A}\,(x_a, \bfk_{\perp a}) 
\nonumber \\ \label{defsiv}
&=& \Delta^N \hf_{a/A^\ua}\,(x_a, k_{\perp a}) \> 
(\hat{\bfp}_A \times \hat{\bfk}_{\perp a}) \cdot \hat{\bfP}^A \> .
\eea
In the last term of the above expression we have explicitly extracted the 
angular dependences, according to the so-called ``Trento conventions'' 
\cite{trento}: $\hat{\bfp}_A$ is the unit vector along the hadron $A$ 
three-momentum, $k_{\perp a} = |\bfk_{\perp a}|$ and $\hat{\bfk}_{\perp a}
= \bfk_{\perp a}/k_{\perp a}$. Parity invariance allows to have a non-zero 
Sivers function only for transverse spin, $\hat{\bfp}_A \cdot \hat{\bfP}^A=0$. 
Often $\Delta^N \hf_{a/\Aup}\,(x_a, k_{\perp a})$ alone is referred to as 
the Sivers function (see Appendix B for related expressions).
For a generic transverse polarization direction 
$\hat{\bfP}^A = (\cos\phi_{S_A}, \, \sin\phi_{S_A}, \, 0)$, one has 
$\Delta \hf_{a/S_A}\,(x_a, \bfk_{\perp a}) = 
\Delta^N \hf_{a/A^\ua}\,(x_a, k_{\perp a}) \, \sin(\phi_{S_A} - \phi_a)$,
where $\phi_a$ is the azimuthal angle (in the hadronic c.m. frame) of 
$\bfk_{\perp a}$.  
 
According to our configuration the hadron transverse polarization is chosen 
along the $+Y$-direction ($\ua$); notice that $Y = Y_{cm}$ for the 
hadron moving along the $+Z_{cm}$-direction, while $Y = - Y_{cm}$ for the 
hadron moving along $-Z_{cm}$, as already noticed after Eq. (2). Then, 
Eq.~(\ref{defsiv}) reads:
\bea
\Delta \hf_{a/S_Y}\,(x_a, \bfk_{\perp a})
&\equiv& \hf _{a/S_Y}\,(x_a, \bfk_{\perp a}) -
\hf _{a/-S_Y}\,(x_a, \bfk_{\perp a}) \nonumber \\ \label{defsivnoi}
&=& \Delta^N \hf_{a/A^\ua}\,(x_a, k_{\perp a}) \> \cos\phi_a \>.
\eea

Similarly, the Boer-Mulders mechanism \cite{amst,dan} (see Appendix B)
allows partons to be transversely polarized inside an unpolarized parent 
hadron. In general, this can be expressed by:
\bea
P^a_j \, \hf_{a/A} &=&
\hf_{s_j/A}^a(x_a, \bfk_{\perp a}) - \hf_{-s_j/A}^a(x_a, \bfk_{\perp a})
\equiv \Delta \hf_{s_j/A}^a(x_a, \bfk_{\perp a})
\nonumber \\ \label{defdan}
&=& \Delta^N \hf_{\aup/A}\,(x_a, k_{\perp a}) \>
(\hat{\bfp}_A \times \hat{\bfk}_{\perp a})_j \>,
\eea
where $P^a_j$ is the $j$-component of the parton polarization in the parton 
helicity frame ($j=x,y,z$). The above equation can also be written as 
\cite{trento} 
\bea
\Delta \hf_{s/A}^a(x_a, \bfk_{\perp a}) &\equiv&
\hf_{s/A}^a(x_a, \bfk_{\perp a}) - \hf_{-s/A}^a(x_a, \bfk_{\perp a})
\nonumber \\ \label{defdang}
&=& \Delta^N \hf_{\aup/A}\,(x_a, k_{\perp a}) \>
(\hat{\bfp}_A \times \hat{\bfk}_{\perp a}) \cdot \hat{\bfP}^a \>,
\eea
where $s$ and $\hat{\bfP}^a$ denote respectively a generic parton spin state 
and the corresponding unit polarization vector, in the parton helicity frame 
(as reached from the parent hadron helicity frame). Notice that, according to 
our configuration, in the hadronic c.m. frame $\hat{\bfy}$ points along the 
$\hat{\bfZ}_{cm} \times \hat{\bfk}_{\perp a}$ direction, Eq.~(\ref{help}). 
It follows that for nucleons moving respectively along the 
$\pm Z_{cm}$-direction one has
\be
\Delta \hf_{s_y/A}^a(x_a, \bfk_{\perp a})
= \pm \Delta^N \hf_{\aup/A}\,(x_a, k_{\perp a}) \>. \label{+-dan}
\ee
It also follows that the analogous function for the $x$-direction is zero, 
$\Delta \hf_{s_x/A}^a = 0$. The function 
$\Delta^N \hf_{\aup/A}\,(x_a, k_{\perp a})$ alone is often referred to 
as the Boer-Mulders function.

Moreover, one can show that the Boer-Mulders function is the same which 
appears in the $y$-component of Eq.~(\ref{DxZ}),
\be
\Delta \hf_{s_y/A}^a = (P^a_y \, \hf_{a/A}) =
(P^a_y \, \hf_{a/A,S_Z}) = \Delta \hf_{{s_y}/+}^a \>, \label{parbm}
\ee
due to parity invariance.

It is worth mentioning that the function 
$(P^a_y \, \hf _{a/S_Y}) = \Delta \hf ^a_{s_y/S_Y}
\equiv \Delta \hf ^a_{s_y/\ua} $
can be decomposed into two terms, the Boer-Mulders 
term which is independent of the hadron transverse polarization, and a term 
which changes sign when the hadron polarization direction is reversed:
\be
\Delta \hf ^a_{s_y/S_Y} = \Delta \hf ^a_{s_y/A} + \Delta ^- \hf ^a_{s_y/S_Y}\,,
\label{del-}
\ee
with
\be
\Delta ^- \hf ^a_{s_y/S_Y} \equiv \frac 12 \, \left[
\Delta \hf ^a_{s_y/\ua} - \Delta \hf ^a_{s_y/\da} \right] =
- \Delta ^- \hf ^a_{s_y/-S_Y} \,. \label{defdel-}
\ee
Notice that 
\be
\Delta \hf^a_{s_x/S_Y} = \Delta^- \hf ^a_{s_x/S_Y}
= - \Delta \hf^a_{s_x/-S_Y} \,. \nonumber
\ee

\subsection{\label{gluon-sect} Gluon polarizations}
Let us now consider the gluon sector (a first study of the unintegrated gluon
distribution functions can be found in Ref.~\cite{mr01}). The helicity density 
matrix for a massless particle with spin $1$ can be written as
\be
\rho_{\lambda_g^{\,}, \lambda^{\prime}_g}^{g/A,S_A}=
\frac{1}{2}\,{\left(
\begin{array}{cc}
1+P_{z}^{g} &
\IoneG-i\ItwoG \\
\IoneG+i \ItwoG & 1-P_{z}^{g}
\end{array}
\right)}_{\!\!\!\!A,S_A} \!\!\!\!\!\!\!
=
\frac{1}{2}\,{\left(
\begin{array}{cc}
1+ P^g_{circ}&
- P^g_{lin} \, e^{-2i\phi}\\
- P^g_{lin} \, e^{2i\phi} & 1-P^g_{circ}
\end{array}
\right)}_{\!\!\!\!A,S_A} \label{rho-gl} \!\!\!, 
\ee 
and we consider it for a gluon $g$ inside the hadron A, in a spin state $S_A$. 
Eq.~(\ref{rho-gl}) refers, in general, to a mixture of circularly and linearly 
polarized states. $P_{circ}^g$ corresponds to $P_{z}^g$, the gluon longitudinal
polarization. The off-diagonal elements of Eq.~(\ref{rho-gl}) are related to 
the linear polarization of the gluons in the $(xy)$ plane at an angle $\phi$ 
to the $x$-axis. The $x$, $y$ and $z$ axes refer to the standard gluon helicity
frame, in which its momentum is $p^\mu = (p,0,0,p)$. $P_{lin}^{g}$ is expressed
in terms of the parameters $\IoneG$ and $\ItwoG$, which are closely related to 
the Stokes parameters used in classical optics; they play a role formally 
analogous to that of the $x$ and $y$-components of the quark polarization 
vector in the quark sector. The use of the parameters $\IoneG$ and $\ItwoG$ 
makes the gluon distribution functions formally similar to those for the 
quarks and considerably simplifies all the formulae for the spin asymmetries 
given in Sections \ref{sigma} and IV.

In analogy to the quark helicity density matrix, Eq.~(\ref{rho-gl}) shows
that:
\bea
&& \rho^g_{++} + \rho^g_{--} = 1 \\
&& \rho^g_{++} - \rho^g_{--} = P^g_z = P^g_{circ} \\
&& 2\,{\rm Re} \rho^g_{-+} = 2\,{\rm Re} \rho^g_{+-} = \IoneG =
-P^g_{lin}\cos(2\phi) \label{Re-rho+-g}\\
&& 2\,{\rm Im} \rho^g_{-+} = -2\,{\rm Im} \rho^g_{+-} = \ItwoG =
 -P^g_{lin}\sin(2\phi)\,.
\label{Im-rho+-g} \eea

As for the quark sector, there are eight independent gluon distribution
functions, which, following Eqs.~(\ref{DxY}-\ref{Dunp}), we label as
\bea
&& (\IoneG \, \hf _{g/A,S_Y})
\equiv
\Delta \hf_{{\Ione}/\ua}^g(x_g, \bfk_{\perp g}) \label{DxY-g}\\
&& (\ItwoG \, \hf _{g/A,S_Y})
\equiv
\Delta \hf_{{\Itwo}/\ua}^g(x_g, \bfk_{\perp g}) \label{DyY-g}\\
&& (P^g_z \, \hf _{g/A,S_Y}) = \Delta \hf_{s_z/S_Y}^g =
\hf _{s_z/\ua}^g -  \hf _{-s_z/\ua}^g \equiv
\Delta \hf_{s_z/\ua}^g(x_g, \bfk_{\perp g}) \label{DzY-g}\\
&& (\IoneG \, \hf _{g/A,S_Z})
\equiv
\Delta \hf_{{\Ione}/+}^g(x_g, \bfk_{\perp g}) \label{DxZ-g}\\
&& (\ItwoG  \, \hf _{g/A,S_Z})
\equiv
\Delta \hf_{{\Itwo}/+}^g(x_g, \bfk_{\perp g}) \label{DyZ-g}\\
&& (P^g_z \, \hf _{g/A,S_Z}) = \Delta \hf_{s_z/S_Z}^g =
\hf _{s_z/+}^g - \hf _{-s_z/+}^g \equiv
\Delta \hf_{s_z/+}^g(x_g, \bfk_{\perp g}) \label{DzZ-g}\\
&& (\hf _{g/A,S_Y}) = \hf_{g/A}(x_g, \bfk_{\perp g}) + \frac{1}{2}\, 
\Delta \hf_{g/S_Y}(x_g,
\bfk_{\perp g})\, \label{Dunp-g}. \label{main-table-g} 
\eea 
Notice that $\Delta \hf_{s_z/+}^g(x_g, \bfk_{\perp g})$ is the usual 
$\bfk_{\perp g}$ dependent gluon helicity distribution function 
$\Delta g(x_g, \bfk_{\perp g})$. The interpretation of 
$\Delta \hf_{{\Ione, \Itwo}/S_A}$ as difference of linearly polarized gluon 
distributions is discussed in the sequel and in Appendix B.

In analogy to Eqs.~(\ref{del-}) and (\ref{defdel-}) we also define a new 
quantity which changes sign when the hadron polarization direction is reversed 
[see Eq.~(\ref{rerho+-g})]:
\be
\Delta \hat f^g_{\Ione/\uparrow} = \Delta \hat f^g_{\Ione/A} +
\Delta^-\hat f^g_{\Ione/\uparrow} \,,
\label{delg-}
\ee
with
\be
\Delta^-\hat f^g_{\Ione/\uparrow} = \frac 12 \, \left[
\Delta \hat f^g_{\Ione/\uparrow} -
\Delta \hat f^g_{\Ione/\downarrow} \right] =
- \Delta^-\hat f^g_{\Ione/\downarrow} \,. \label{defdelg-}
\ee

Although gluons cannot carry any transverse spin, there is a strong analogy 
between transversely polarized quarks and linearly polarized gluons; for 
example, analogous to the Boer--Mulders case for quarks, it is possible to 
have a linearly polarized gluon inside an unpolarized nucleon, corresponding 
to a non-vanishing $\Ione^g \hat f_{g/A} = \Delta\hat f_{\Ione/A}^g$. This 
mechanism has never been explored before. Its structure is linked to the
spin 1 Cartesian tensor $T_{ij}$ (see, {\it e.g.}, Section 3.1.12 of 
Ref.~\cite{elliot}), which is symmetric and traceless. 
For a massless particle one 
has:
\bea
&& \quad\quad\quad\quad\quad\quad\quad\quad\quad
T_{zz} = \frac{1}{\sqrt 6} \label{tija} \\
&& \Ione = \sqrt{\frac 23} \left( T_{xx} - T_{yy} \right) \label{tijb}
\quad\quad\quad\quad
\Itwo = 2 \, \sqrt{\frac 23} \, T_{xy} \>. \label{tijc}
\eea
Because of (\ref{tija}), the traceless condition and parity invariance,
it is only possible to construct one scalar structure that depends
non-trivially on the $T_{ij}$.

Using the three-vectors at our disposal -- the gluon momentum $\bfp$, its
transverse momentum $\bfk_\perp$ and the parent hadron momentum $\bfp_A$ --
we define:
\be 
\hat{\bfu} = \frac{\hat{\bfk}_\perp - (\hat{\bfk}_\perp \cdot \hat{\bfp}) \, 
\hat{\bfp}} {\hat{\bfp} \cdot \hat{\bfp}_A} \quad\quad\quad\quad\quad 
\hat{\bfv} = \hat{\bfp}_A \times \hat{\bfk}_\perp \label{uv} 
\ee
and introduce a tensor {\bf T} whose components are $T_{ij}$. The
only possible structure is then:
\bea 
T_{ij} \, \hat f_{g/A}(x, k_\perp) &=& \sqrt{\frac{3}{2}} \, \left[ 
\frac{1}{2} \, \Delta^N \hat f^g_{\Ione/A}(x, \kt) 
\left( \hat{u}_i \hat{u}_j - \hat{v}_i \hat{v}_j \right) \right. \nonumber \\ 
&-& \left. \frac{1}{6} \, \hat f_{g/A}(x, k_\perp) \,
\left( \hat{u}_i \hat{u}_j + \hat{v}_i \hat{v}_j - 2 \, \hat{p}_i \hat{p}_j
\right) \right] \>, 
\eea
which is the gluon tensorial analogue of Eq.~(\ref{defdan}). When nucleon 
$A$ moves along or opposite the $Z_{cm}$-axis this reduces to:
\be 
\Ione^g \hat f_{g/A}(x, \bfk_\perp) = \Delta \hat 
f^g_{\Ione/A}(x, \bfk_\perp) = \Delta^N
\hat f^g_{\Ione/A}(x, \kt) \label{+-dang} 
\ee
in analogy to Eq.~(\ref{+-dan}). Notice that, in this case, there 
is no $\pm$ sign on the r.h.s. of Eq.~(\ref{+-dang}).

One can also show that the linear polarization $\IoneG$ is independent of
any longitudinal polarization of the nucleon, {\it i.e.}:
\be
\Delta \hat f^g_{\Ione/A} = \Delta \hat f^g_{\Ione/A,S_Z} =
\Delta \hat f^g_{\Ione/+} \>,
\ee
as in Eq.~(\ref{parbm}).

\subsection{\label{frag} Quark and gluon fragmentation functions into 
unpolarized hadrons}

As already mentioned in Section \ref{Intro},
for the fragmentation process in general we define
 \be
 \hat D^{\la^{\,}_C,\la^{\prime}_C}_{\la^{\,}_c,\la^{\prime}_c}
 (z, \bfk_{\perp C})
 = \> \sumint_{X, \la_{X}}
 {\hat{\cal D}}_{\la^{\,}_C,\la^{\,}_{X};\la^{\,}_c}(z, \bfk_{\perp C}) \,
 {\hat{\cal D}}^*_{\la^{\prime}_C,\la^{\,}_{X}; \la^{\prime}_c}
 (z, \bfk_{\perp C})
 \, . \label{framp}
 \ee
The analogous quantity for parton distributions can be found in 
Eq.~(\ref{defFF}). $\hat{{\cal D}}_{\la^{\,}_C,\la^{\,}_{X}; \la^{\,}_c}$ 
is the {\it fragmentation amplitude} describing the process $c\to C+X$ in which
the parton $c$ from the elementary scattering $ab\to cd$ generates the 
detected final hadron $C$, with light-cone momentum fraction $z$ and 
transverse momentum $\bfk _{\perp C}$. If we denote by $\phi_C^H$ the 
azimuthal angle of the hadron $C$ in the parton $c$ helicity frame, we have
 \be
 \hat{\cal D}_{\la^{\,}_C,\la^{\,}_{X}; \la^{\,}_c}(z, \bfk_{\perp C}) =
 {\cal D}_{\la^{\,}_C,\la^{\,}_{X}; \la^{\,}_c}(z, k_{\perp C}) \>
 e^{i\la^{\,}_c \phi_C^H} \>, \label{fragphi}
 \ee
similarly to Eq.~(\ref{dampphi}) for parton distribution amplitudes.
Eqs.~(\ref{framp}) and (\ref{fragphi}) then give the generalized
fragmentation function
\be
{\hat D}_{\la^{\,}_{c},\la^{\prime}_c}^{\la^{\,}_C,\la^{\prime}_C}
(z, \bfk_{\perp C})
={D}_{\la^{\,}_{c},\la^{\prime}_c}^{\la^{\,}_C,\la^{\prime}_C}
(z, k_{\perp C}) \>
e^{i(\la^{\,}_{c} - \la^{\prime}_c)\phi_C^H} \>,
\ee
while the corresponding generalized distribution function is given in
Eq.~(\ref{fft-ff}).

If hadron $C$ is unpolarized, the
generalized fragmentation function ${\hat D}$ simply becomes
\be
{\hat D}_{\la^{\,}_{c},\la^{\prime}_c}^{C/c}(z, \bfk_{\perp C})
= \sum_{\la^{\,}_{C}}
\hat D^{\la^{\,}_C,\la^{\,}_C}_{\la^{\,}_c,\la^{\prime}_c}
(z, \bfk_{\perp C}) = {D}_{\la^{\,}_{c},\la^{\prime}_c}^{C/c} (z, k_{\perp C}) 
\> e^{i(\la^{\,}_{c} - \la^{\prime}_c)\phi_C^H} \>,
\label{ddt-dd}
\ee
and fulfills the parity properties given by
\be
{D}_{-\la^{\,}_{c},-\la^{\prime}_c}^{C/c}(z, k_{\perp C})
= (-1)^{2s_c} \> (-1)^{\la^{\,}_c + \la^{\prime}_c} \>
{D}_{\la^{\,}_{c},\la^{\prime}_c}^{C/c}(z, k_{\perp C}) \,.
\label{parDD}
\ee
If parton $c$ is a quark, $s_c=1/2$ and the helicities
$\la^{\,}_{c}$ and $\la^{\prime}_c$ will be either $=+1/2$ or $-1/2$,
whereas if parton $c$ is a gluon, $s_c=1$ and
$\la^{\,}_{c}$ and $\la^{\prime}_c$ will be either $=+1$ or $-1$.

\goodbreak
\vskip 6pt \noindent
{\it Quark fragmentation functions} 
\vskip 4pt
\nobreak

For quarks, from Eqs.~(\ref{framp}), (\ref{ddt-dd}) and (\ref{parDD}) we 
obtain the following relations
\bea
&&{\hat D}_{++}^{C/q}(z,\bfk _{\perp C}) =
D_{++}^{C/q}(z,k _{\perp C}) \equiv
{\hat D}_{C/q}(z, k _{\perp C}) \nonumber \\
&&
{\hat D}_{--}^{C/q}(z,\bfk _{\perp C}) =
{\hat D}_{++}^{C/q}(z,\bfk _{\perp C}) \label{parqe}
\eea
for equal helicity indices, and
\bea
&&{\hat D}_{+-}^{C/q}(z,\bfk _{\perp C}) =
D_{+-}^{C/q}(z,k _{\perp C}) \,e^{i\phi_C^H} =
-D_{-+}^{C/q}(z,k _{\perp C}) \,e^{i\phi_C^H} \nonumber \\
&&{\hat D}_{-+}^{C/q}(z,\bfk _{\perp C}) =
D_{-+}^{C/q}(z,k _{\perp C}) \,e^{-i\phi_C^H} =
-D_{+-}^{C/q}(z,k _{\perp C}) \,e^{-i\phi_C^H} \label{D+-q} \label{parqu}\\
&&{[D_{+-}^{C/q}(z,k _{\perp C})]}^* = - D_{+-}^{C/q}(z,k _{\perp C})
\nonumber
\eea
for unequal helicity indices.

${\hat D}_{C/q}(z,k _{\perp C})$ is the $k_{\perp C}$ dependent fragmentation 
function describing the hadronization of an unpolarized quark $q$ into an 
unpolarized hadron $C$. Notice that it does not actually depend on the 
direction of $\bfk _{\perp C}$, but only on its modulus. When integrated 
over the intrinsic transverse momentum, this function gives us the usual 
unpolarized fragmentation function $D_{C/c}(z)$, see Eq.~(\ref{fr}), 
\be 
D_{C/q}(z) = \frac{1}{2} \sum_{\la^{\,}_q} \int d^2\bfk_{\perp C} \, \hat
D^{C/q}_{\la^{\,}_q,\la^{\,}_q}(z, \bfk_{\perp C}) \,.  
\ee 

Eqs.~(\ref{D+-q}) tell us that the fragmentation function 
$D_{+-}^{C/q}(z,k _{\perp C})$ is an independent purely imaginary quantity. 
It is related to the Collins quark fragmentation function by the following
expression: 
\be 
-2i\,D_{+-}^{C/q}(z,k _{\perp C}) = 2\,{\rm Im}\,D_{+-}^{C/q}(z,k _{\perp C}) 
\equiv \Delta^N \hat{D}_{C/\qup} (z,k _{\perp C}) \label{collins}\,, 
\ee 
and gives the difference between the number densities of unpolarized hadrons 
$C$ resulting from the fragmentation of a quark $q$ polarized along the 
$+y$-direction and a quark polarized along the $-y$-direction, in the quark 
helicity  frame in which the fragmentation process occurs in the $(xz)$ plane.
In general one has, analogously to Eq.~(\ref{defsiv}) for the Sivers function,
\bea
\Delta \hat D_{C/q,s}\,(z, \bfk_{\perp C})
&=& \hat D_{C/q,s}\,(z, \bfk_{\perp C}) -
\hat D_{C/q,-s}\,(z, \bfk_{\perp C})
\nonumber \\ \label{defcol}
&=& \Delta^N \hat D_{C/\qup}\,(z, k_{\perp C}) \>
(\hat{\bfp}_q \times \hat{\bfk}_{\perp C}) \cdot \hat{\bfP}^q \>.
\eea
If $\hat{\bfP}^q$ points along the $\hat{\bfy}$-direction Eq.~(\ref{defcol}) 
reads, in analogy to Eq.~(\ref{defsivnoi}): 
\bea
\Delta \hat D_{C/q,s_y}\,(z, \bfk_{\perp C})
&=& \hat D_{C/q,s_y}\,(z, \bfk_{\perp C}) -
\hat D_{C/q,-s_y}\,(z, \bfk_{\perp C})
\nonumber \\ \label{defcolnoi}
&=& \Delta^N \hat D_{C/\qup}\,(z, k_{\perp C}) \> \cos\phi_C^H \>, 
\eea
consistently with Eq.~(\ref{ddt-dd}). The explicit expression of $\phi_C^H$ 
in terms of the overall hadronic variables, in the $AB$ c.m. frame, can be 
found in Eq. (45) of Ref. \cite{noi} and in Appendix A, Eq.~(A28).
\goodbreak
\vskip 6pt \noindent
{\it Gluon fragmentation functions} 
\vskip 4pt
\nobreak

The gluon fragmentation functions with equal helicity indices obey the same
parity rules (\ref{parqe}) as the quark ones:
\bea
&&{\hat D}_{++}^{C/g}(z,\bfk _{\perp C}) =
D_{++}^{C/g}(z,k _{\perp C}) \equiv
{\hat D}_{C/g}(z,k _{\perp C}) \nonumber \\
&& {\hat D}_{--}^{C/g}(z,\bfk _{\perp C}) =
{\hat D}_{++}^{C/g}(z,\bfk _{\perp C}) \,;
\eea
however, as implied by Eq.~(\ref{parDD}), a different sign, with respect
to the quark case (\ref{parqu}), appears in the parity relations for the
generalized gluon fragmentation functions with unequal helicity indices:
\bea
&&{\hat D}_{+-}^{C/g}(z,\bfk _{\perp C}) =
D_{+-}^{C/g}(z,k _{\perp C}) \,e^{2i\phi_C^H} =
D_{-+}^{C/g}(z,k _{\perp C}) \,e^{2i\phi_C^H} \nonumber \\
&&{\hat D}_{-+}^{C/g}(z,\bfk _{\perp C}) =
D_{-+}^{C/g}(z,k _{\perp C}) \,e^{-2i\phi_C^H} =
D_{+-}^{C/g}(z,k _{\perp C}) \,e^{-2i\phi_C^H} \label{D+-g}\\
&&{[D_{+-}^{C/g}(z,k _{\perp C})]}^* =
D_{+-}^{C/g}(z,k _{\perp C})\,.
\nonumber
\eea
The above equations show that $D_{+-}^{C/g}(z,k _{\perp C})$ is an
independent, real quantity. Notice that the gluon Collins fragmentation
function cannot exist, since there is not such an object as a transversely
spin polarized real gluon. However, similarly to what happens for the gluon 
parton distributions, the fragmentation function $D^{C/g}_{+-}(z,k _{\perp C})$
is related to the fragmentation process into a spinless hadron $C$ of a
linearly polarized gluon. In analogy to Eq.~(\ref{collins}) we have 
\be 
2\,D_{+-}^{C/g}(z,k _{\perp C}) = 2\,{\rm Re}\,D_{+-}^{C/g}(z,k _{\perp C}) 
\equiv \Delta^N \hat{D}_{C/\IoneG} (z,k _{\perp C}) \label{collinslike}\,, 
\ee 
which gives the difference between the number densities of unpolarized hadrons 
$C$ resulting from the fragmentation of a gluon linearly polarized along the 
$x$-direction and a gluon linearly polarized along the $y$-direction, in the 
gluon helicity frame in which the fragmentation process occurs in the 
$(xz)$ plane.

\section{\label{sigma} Kernels}

As we can see from Eq.~(\ref{gen1}), the computation of the cross section 
corresponding to any polarized hadronic process $(A,S_A) + (B,S_B) \to C + X$ 
requires the evaluation and integration, for each elementary process 
$a+b\to c+d$, of the general kernel 
\bea
\Sigma(S_A,S_B)^{ab\to cd} &=& \sum _{\{\lambda\}} \rho_{\la^{\,}_a,
\la^{\prime}_a}^{a/A,S_A} \, \hat f_{a/A,S_A}(x_a,\bfk_{\perp a}) \> 
\rho_{\la^{\,}_b,
\la^{\prime}_b}^{b/B,S_B} \, \hat f_{b/B,S_B}(x_b,\bfk_{\perp b})
\label{kern}  \\
&\times& \hat M_{\la^{\,}_c, \la^{\,}_d; \la^{\,}_a, \la^{\,}_b} \, 
\hat M^*_{\la^{\prime}_c,
\la^{\,}_d; \la^{\prime}_a, \la^{\prime}_b} \> \hat
D^{\la^{\,}_C,\la^{\,}_C}_{\la^{\,}_c,\la^{\prime}_c}(z,\bfk_{\perp C})\,.
\nonumber 
\eea

Whereas the hadronic process $(A,S_A) + (B,S_B) \to C + X$ takes place, 
according to our choice, in the $(XZ)_{cm}$ plane, all the elementary 
processes involved, $A(B) \to a(b) + X$, $ab \to cd$ and $c \to C + X$ do not, 
since all parton and hadron momenta, $\bfp_a, \, \bfp_b, \, \bfp_C$  have 
transverse components $\bfk_{\perp a}, \, \bfk_{\perp b}, \, \bfk_{\perp C}$.
This ``out of $(XZ)_{cm}$ plane'' geometry induces in the fragmentation 
process the phase given in Eq.~(\ref{ddt-dd}) and, in the distribution 
functions, the phase appearing in Eq.~(\ref{fft-ff}).

Analogously, the elementary QCD process $ab \to cd$, whose helicity
amplitudes are well known in the $ab$ center of mass frame, is not, in
general, a planar process anymore when observed from the $AB$ center of mass
frame, the laboratory frame, where we are performing our computations.
However, we can go from the actual $\bfp_a \, \bfp_b \to \bfp_c \, \bfp_d$
configuration, as seen in the laboratory frame, to the canonical one in
which the $ab \to cd$ process takes place in the $ab$ c.m. frame and in
the $(XZ)_{cm}$ plane, by performing one boost and appropriate 
rotations, as described in full detail in Ref.~\cite{noi}. These
transformations introduce some highly non trivial phases in the helicity
amplitudes $\hat M_{\la^{\,}_c, \la^{\,}_d; \la^{\,}_a, \la^{\,}_b}$, which
are the direct consequence of the complicated non planar kinematics.
The relation between these amplitudes (which we need in our computations)
and the usual, canonical amplitudes ${\hat M}^0$, defined in the partonic
$ab\to cd$ c.m. frame, is the following \cite{noi}
\be
\hat M_{\la^{\,}_c, \la^{\,}_d; \la^{\,}_a, \la^{\,}_b} \!
= \hat M^0
_{\la^{\,}_c, \la^{\,}_d; \la^{\,}_a, \la^{\,}_b}
\, e^{-i (\la^{\,}_a \xi _a + \la^{\,}_b \xi _b -
          \la^{\,}_c \xi _c - \la^{\,}_d \xi _d)}
\, e^{-i [(\la^{\,}_a - \la^{\,}_b) \tilde \xi _a -
         (\la^{\,}_c - \la^{\,}_d) \tilde \xi _c]}
\, e^{i(\la^{\,}_a - \la^{\,}_b)\phi^{\prime\prime}_c}
\label{M-M0}
\ee
with $\xi _j$, $\tilde \xi _j$ ($j=a,b,c,d$) and  $\phi^{\prime\prime}_c$
defined in Eqs.~(35-42) of Ref.~\cite{noi} and in Appendix A. The parity
properties of the canonical c.m. amplitudes $\hat M^0$ are the usual ones:
\be
\hat M^0_{-\la^{\,}_c, -\la^{\,}_d; -\la^{\,}_a, -\la^{\,}_b} =
\eta_a \, \eta_b \, \eta_c \, \eta_d \, (-1)^{s_a + s_b - s_c - s_d} \>
(-1)^{(\la^{\,}_a - \la^{\,}_b) - (\la^{\,}_c - \la^{\,}_d)} \, 
\hat M^0_{\la^{\,}_c, \la^{\,}_d; \la^{\,}_a, \la^{\,}_b} \>,
\label{parM0}
\ee
where $\eta_i$ is the intrinsic parity factor for particle $i$. For massless
partons there are only three independent elementary amplitudes $\hat M^0$
corresponding to the $ab\to cd$ processes we are interested in; this
allows us to adopt the following notation
\bea
\hat M_{++;++} & \equiv & \hat M_1^0 \,e^{i\varphi_1} \nonumber \\
\hat M_{-+;-+} & \equiv & \hat M_2^0 \,e^{i\varphi_2} \label{phases} \\
\hat M_{-+;+-} & \equiv & \hat M_3^0 \,e^{i\varphi_3}, \nonumber
\eea
where  $\hat M_1^0$, $\hat M_2^0$ and $\hat M_3^0$ are defined as
\be
\hat M^0_{+,+;+,+} \equiv \hat M_1^0 \quad\quad\quad
\hat M^0_{-,+;-,+} \equiv \hat M_2^0 \quad\quad\quad
\hat M^0_{-,+;+,-} \equiv \hat M_3^0 \>, \label{Mqq}
\ee
and the phases $\varphi_1$, $\varphi_2$ and $\varphi_3$ are given by replacing
in Eq.~(\ref{M-M0}) the appropriate value for the helicities $\lambda_i$,
$i=a,b,c,d$. Indeed, the $+$ and $-$ subscripts refer to $(+1/2)$ and 
$(-1/2)$ helicities for quarks, and to $(+1)$ and $(-1)$ helicities for 
gluons. 

All other amplitudes are obtained from Eqs.~(\ref{M-M0}), (\ref{phases}) and 
(\ref{Mqq}), exploiting the parity properties (\ref{parM0}); notice that the 
presence of the phases $\varphi_j$ implies that the parity relations for 
the amplitudes $\hat M$ are not as simple as those for the ${\hat M}^0$.
\mbox{}From Lorentz and rotational invariance properties \cite{elliot} one
can obtain the following useful expressions relating the canonical amplitudes
for processes which only differ by the exchange of the two initial
partons, $a \leftrightarrow b $, or of the two final partons,
$c \leftrightarrow d $:
\bea
&&
\hat M_{\la^{\,}_c, \la^{\,}_d; \la^{\,}_b, \la^{\,}_a}^{0, \, ba \to cd}
(\theta) =
\hat M_{\la^{\,}_c, \la^{\,}_d; \la^{\,}_a, \la^{\,}_b}^{0, \, ab \to cd}
(\pi - \theta) \, e^{-i\pi(\la^{\,}_c - \la^{\,}_d)}
\label{exchange-ab} \\ &&
\hat M_{\la^{\,}_d, \la^{\,}_c; \la^{\,}_a, \la^{\,}_b}^{0, \, ab \to dc}
(\theta) =
\hat M_{\la^{\,}_c, \la^{\,}_d; \la^{\,}_a, \la^{\,}_b}^{0, \, ab \to cd}
(\pi -\theta) \, e^{-i \pi(\la^{\,}_a - \la^{\,}_b)}\,,
\label{exchange-cd}
\eea
where the scattering angle $\theta$ is defined in the canonical partonic
c.m. frame. To be precise, the above relationships hold up to an overall,
helicity independent, phase; since only bilinear combinations of the
amplitudes occurr in the expressions for physical observables, we fix
such a phase to be $+1$.

There are eight elementary contributions $ab\to cd$ which we have to consider
separately
\bea
&& q_a q_b \to q_c q_d \,, \quad  g_a g_b \to g_c g_d \,, \nonumber \\
&& q g \to q g \,, \quad  g q \to g q \,,\nonumber \\
&& q g \to g q \,, \quad  g q \to q g \,, \\
&& g_a g_b \to q \bar{q}\,, \quad  q \bar{q} \to g_c g_d \,, \nonumber
\eea
where $q$ can in general be either a quark or an antiquark. The subscripts
$a,b,c,d$ for quarks, when necessary, identify the flavour (only in processes
where different flavours can be present); for gluons, these labels identify
the corresponding hadron ($a \to A, \> b \to B, \> c \to C$). By performing
the explicit sums in Eq.~(\ref{kern}), we obtain the kernels for each of the 
elementary processes. Note that the new aspect of our calculation is the 
oppearance of the phases which is a reflection of the noncollinear 
kinematics. For convenience we also give explicit expressions for the
combination of the partonic c.m. amplitudes $\hat M_j^0$ which are needed. 
\vskip 6pt
\noindent {\bf 1)} $q_a q_b \to q_c q_d$ {\bf processes}
\bea 
&& \Sigma(S_A,S_B)^{q_a q_b \to q_c q_d} = \frac 12 \, 
\hat D_{C/c}(z, k_{\perp C}) \>
\hat f_{a/S_A}(x_a, \bfk_{\perp a}) \> \hat f_{b/S_B}(x_b, \bfk_{\perp b})
\times \nonumber \\
&& \quad\quad\quad\quad\quad
\Biggl\{ \left( |\hat M^0_1|^2 + |\hat M^0_2|^2 + |\hat M^0_3|^2 \right)
+ P_z^a \, P_z^b \left( |\hat M^0_1|^2 - |\hat M^0_2|^2 - |\hat M^0_3|^2
\right) \nonumber \\
&& \quad\quad\quad\quad\quad
+ \>  2 \hat M^0_2 \, \hat M^0_3 \left[ \left(
P_x^a \, P_x^b + P_y^a \, P_y^b \right) \, \cos(\vphi_{3}-\vphi_{2}) - \left(
P_x^a \, P_y^b - P_y^a \, P_x^b \right) \, \sin(\vphi_{3}-\vphi_{2})
\right] \Biggr\} \nonumber \\
&& \quad\quad\quad\quad\quad - \> \frac 12 \, 
\Delta^N\hat D_{C/\cupar}(z, k_{\perp C}) \>
\hat f_{a/S_A}(x_a, \bfk_{\perp a}) \> \hat f_{b/S_B}(x_b, \bfk_{\perp b}) 
\times \label{qqqq} \\
&& \quad\quad\quad\quad\quad
\Biggl\{ \hat M^0_1 \, \hat M^0_2 \, \left[
  P_x^a \, \sin(\vphi_1 - \vphi_2 + \phi_C^H)
- P_y^a \, \cos(\vphi_1 - \vphi_2 + \phi_C^H) \right] \nonumber \\
&& \quad\quad\quad\quad\quad
+ \> \hat M^0_1 \, \hat M^0_3 \, \left[
  P_x^b \, \sin(\vphi_1 - \vphi_3 + \phi_C^H)
- P_y^b \, \cos(\vphi_1 - \vphi_3 + \phi_C^H) \right] \Biggr\} \>, \nonumber
\eea
where (including color factors)
\bea
&& |\hat M_{1}^{0}|^{2} = \frac{8}{9} \, g_s^4 \left[
\frac{\hat{s}^{2}}{\hat{t}^{2}} + \delta_{ab} \, \left(
\frac{\hat{s}^{2}}{\hat{u}^{2}}
- \frac{2}{3} \, \frac{\hat{s}^{2}}{\hat t\hat u} \right) \right]
\quad\quad\quad
|\hat M_{2}^{0}|^{2} = \frac{8}{9} \, g_s^4 \,
\frac{\hat{u}^{2}}{\hat{t}^{2}} \nonumber \\
&& |\hat M_{3}^{0}|^{2} = \delta_{ab} \, \frac{8}{9} \, g_s^4 \,
\frac{\hat{t}^{2}}{\hat{u}^{2}}
\quad\quad\quad
\hat M_{1}^{0} \, \hat M_{2}^{0} = \frac{8}{9} \, g_s^4
\left(-\frac{\hat s\hat u}{\hat{t}^{2}} + \delta_{ab} \, \frac{1}{3} \,
\frac{\hat s}{\hat t}\right) \\
&& \hat M_{1}^{0} \, \hat M_{3}^{0} = \delta_{ab} \, \frac{8}{9} \, g_s^4
\left( \frac{\hat s\hat t}{\hat{u}^{2}} - \frac{1}{3} \,
\frac{\hat s}{\hat u}\right) \quad\quad\quad
\hat M_{2}^{0} \, \hat M_{3}^{0} = \delta_{ab} \, \frac{8}{27} \, g_s^4
\nonumber
\eea
if $q_a$, $q_b$, $q_c$ and $q_d$ are either all quarks or all anti-quarks, and
\bea
 &&|\hat M_{1}^{0}|^{2} = \delta_{ac} \, \frac{8}{9} \, g_s^4 \,
 \frac{\hat{s}^{2}}{\hat{t}^{2}}
 \quad\quad\quad |\hat M_{2}^{0}|^{2} = \frac{8}{9} \,g_s^4
 \left(\delta_{ab} \, \frac{\hat{u}^{2}}{\hat{s}^{2}} + \delta_{ac} \,
 \frac{{\hat u}^{2}}{\hat{t}^{2}} - \delta_{ab} \, \delta_{ac} \,
 \frac{2}{3}\,\frac{\hat{u}^{2}}{\hat s\hat t}\right) \nonumber\\
 && |\hat M_{3}^{0}|^{2} = \delta_{ab} \, \frac{8}{9} \, g_s^4
 \frac{\hat{t}^{2}}{\hat{s}^{2}} \quad\quad\quad
 \hat M_{1}^{0} \, \hat M_{2}^{0} = \frac{8}{9} \, g_s^4 \, \delta_{ac}
 \left(-\frac{\hat s\hat u}{\hat{t}^{2}} + \delta_{ab} \, \frac{1}{3} \,
 \frac{\hat u}{\hat t}\right) \\
 && \hat M_{1}^{0} \, \hat M_{3}^{0} = \delta_{ab}\, \delta_{ac} \, 
 \frac{8}{27} \, g_s^4
 \quad\quad\quad \hat M_{2}^{0} \, \hat M_{3}^{0} = \frac{8}{9}\,g_s^4 \, 
 \delta_{ab}
 \left(\frac{\hat u\hat t}{\hat{s}^{2}}-\delta_{ac} \frac{1}{3}
 \frac{\hat u}{\hat s}\right)
 \nonumber
 \eea
for any combination of the type $q_a{\bar q}_b \to q_c{\bar q}_d$.

\vskip 6pt

\noindent {\bf 2)} $qg \to qg$ {\bf processes}
\bea && \Sigma(S_A,S_B)^{q g \to q g} = \frac 12 \,
 \hat D_{C/q}(z, k_{\perp C}) \>
\hat f_{q/S_A}(x_a, \bfk_{\perp a}) \> \hat f_{g/S_B}(x_b, \bfk_{\perp b})
\times \nonumber \\
&& \hskip 60pt \Biggl\{ \left( |\hat M^0_1|^2 + |\hat M^0_2|^2 \right)
+ P_z^q \, P_z^g \left( |\hat M^0_1|^2 - |\hat M^0_2|^2 \right)
\Biggr\}\nonumber \\
&& \hskip 60pt  - \> \frac 12 \, \Delta^N\hat D_{C/\qup}(z, k_{\perp C}) \>
 \hat f_{q/S_A}(x_a, \bfk_{\perp a}) \> \hat f_{g/S_B}(x_b, \bfk_{\perp b}) 
\times \label{qgqg}\\
&& \hskip 60pt \Biggl\{ \hat M^0_1 \, \hat M^0_2
\left[ P_x^q \, \sin(\vphi_1 - \vphi_2 + \phi_C^H)
- P_y^q \, \cos(\vphi_1 - \vphi_2 + \phi_C^H) \right] \Biggr\} \>. \nonumber
\eea
In this case the amplitude ${\hat M}_3^0$ is zero because it violates 
helicity conservation, and
\be
|\hat M_{1}^{0}|^{2}= \frac{8}{9}\,g_s^4
\left(-\frac{\hat s}{\hat u}+\frac{9}{4}\frac{\hat{s}^{2}}{\hat{t}^{2}}\right)
\quad\> | \hat M_{2}^{0}|^{2}=\frac{8}{9}\,g_s^4
\left(-\frac{\hat u}{\hat s}+\frac{9}{4}\frac{\hat{u}^{2}}{\hat{t}^{2}}\right)
\quad\> \hat M_{1}^{0} \hat M_{2}^{0}=\frac{8}{9}\,g_s^4\left(-1+\frac{9}{4}
\frac{\hat u \hat s}{\hat{t}^{2}}\right). 
\ee
\vskip 6pt
 \noindent {\bf 3)} $gq \to qg$ {\bf processes}
 \bea
 && \Sigma(S_A,S_B)^{g q \to q g} =
 \frac 12 \, \hat D_{C/q}(z, k_{\perp C}) \>
 \hat f_{g/S_A}(x_a, \bfk_{\perp a}) \> \hat f_{q/S_B}(x_b, \bfk_{\perp b})
 \times \nonumber \\
 && \hskip 60pt \Biggl\{ \left( |\hat M^0_1|^2 + |\hat M^0_3|^2 \right)
 + P_z^g \, P_z^q \left( |\hat M^0_1|^2 - |\hat M^0_3|^2 \right)
 \Biggr\}\nonumber \\
 && \hskip 60pt  - \> \frac 12 \, \Delta^N\hat D_{C/\qup}(z, k_{\perp C}) \>
 \hat f_{g/S_A}(x_a, \bfk_{\perp a}) \> \hat f_{q/S_B}(x_b, \bfk_{\perp b}) 
 \times \label{gqqg}\\
 && \hskip 60pt \Biggl\{ \hat M^0_1 \, \hat M^0_3
 \left[ P_x^q \, \sin(\vphi_1 - \vphi_3 + \phi_C^H)
 - P_y^q \, \cos(\vphi_1 - \vphi_3 + \phi_C^H) \right] \Biggr\} \,, \nonumber
 \eea
where now the amplitude ${\hat M}_2^0$ is zero because of QCD
helicity conservation and the amplitudes $[{\hat M}_1^0]_{gq \to qg}$ and
$[{\hat M}_3^0]_{gq \to qg}$ can be obtained from
$[{\hat M}_1^0]_{qg \to qg}$ and $[{\hat M}_2^0]_{qg \to qg}$ by applying
Eq.~(\ref{exchange-ab}).

\vskip 6pt
\noindent {\bf 4)} $qg \to gq$ {\bf processes}
 \bea
 && \Sigma(S_A,S_B)^{q g \to g q} =
 \frac 12 \, \hat D_{C/g}(z, k_{\perp C}) \>
 \hat f_{q/S_A}(x_a, \bfk_{\perp a}) \> \hat f_{g/S_B}(x_b, \bfk_{\perp b})
 \times \nonumber \\
 && \hskip 60pt \Biggl\{ \left( |\hat M^0_1|^2 + |\hat M^0_3|^2 \right)
 + P_z^q \, P_z^g \left( |\hat M^0_1|^2 - |\hat M^0_3|^2 \right)
 \Biggr\}\nonumber \\
 && \hskip 60pt  + \> \frac 12 \, \Delta^N \hat{D}_{C/\IoneG}(z, k_{\perp C}) 
 \> \hat f_{q/S_A}(x_a, \bfk_{\perp a}) \> \hat f_{g/S_B}(x_b, \bfk_{\perp b}) 
 \times \label{qggq}\\
 && \hskip 60pt \Biggl\{ \hat M^0_1 \, \hat M^0_3
 \left[ \Ione^g \, \cos(\vphi_1 - \vphi_3 + 2\phi_C^H)
 + \Itwo^g \, \sin(\vphi_1 - \vphi_3 + 2\phi_C^H) \right] \Biggr\} \,, 
 \nonumber
 \eea
where again the amplitude ${\hat M}_2^0$ is zero  because of
helicity conservation and the amplitudes $[{\hat M_1}^0]_{qg \to gq}$ and
$[{\hat M}_3^0]_{qg \to gq}$ can be obtained from $[{\hat M}_1^0]_{qg \to qg}$
and $[{\hat M}_2^0]_{qg \to qg}$ by applying Eq.~(\ref{exchange-cd}).

\vskip 6pt
\noindent {\bf 5)} $gq \to gq$ {\bf processes}
 \bea
 && \Sigma(S_A,S_B)^{g q \to g q} =
 \frac 12 \, \hat D_{C/g}(z, k_{\perp C}) \>
 \hat f_{g/S_A}(x_a, \bfk_{\perp a}) \> \hat f_{q/S_B}(x_b, \bfk_{\perp b})
 \times \nonumber \\
 && \hskip 60pt \Biggl\{ \left( |\hat M^0_1|^2 + |\hat M^0_2|^2 \right)
 + P_z^g \, P_z^q \left( |\hat M^0_1|^2 - |\hat M^0_2|^2 \right)
 \Biggr\}\nonumber \\
 && \hskip 60pt  + \>  \frac 12 \, \Delta^N \hat{D}_{C/\IoneG}(z, k_{\perp C}) 
 \> \hat f_{g/S_A}(x_a, \bfk_{\perp a}) \> \hat f_{q/S_B}(x_b, \bfk_{\perp b}) 
 \times \label{gqgq}\\
 && \hskip 60pt \Biggl\{ \hat M^0_1 \, \hat M^0_2
 \left[ \Ione^g \, \cos(\vphi_1 - \vphi_2 + 2\phi_C^H)
 + \Itwo^g \, \sin(\vphi_1 - \vphi_2 + 2\phi_C^H) \right] \Biggr\} \>,\nonumber
 \eea
where the amplitude ${\hat M}_3^0$ is zero  because it violates
helicity conservation and the amplitudes $[{\hat M}_1^0]_{gq \to gq}$ and
$[{\hat M}_2^0]_{gq \to gq}$ can be obtained from
$[{\hat M}_1^0]_{gq \to qg}$ and
$[{\hat M}_3^0]_{gq \to qg}$ by applying Eq.~(\ref{exchange-cd}).
\vskip 6pt
\noindent {\bf 6)} $q\bar q \to g_c g_d$ {\bf processes}
 \bea
 && \hskip -20pt \Sigma(S_A,S_B)^{q \bar q \to g_c g_d} = \frac 12 \, 
 \hat D_{C/g}(z,k_{\perp C}) \> \hat f_{q/S_A}(x_a, \bfk_{\perp a}) \>
 \hat f_{\bar q/S_B}(x_b, \bfk_{\perp b}) \times \nonumber \\
 && \hskip 40pt \Biggl\{ \left( 1 - P_z^q \, P_z^{\bar q} \right)
 \left( |\hat M^0_2|^2 + |\hat M^0_3|^2 \right) \label{qqgg} \\
 && \hskip 40 pt + \> 2  \hat M^0_2 \, \hat M^0_3 \left[
 \left( P_x^q \, P_x^{\bar q} + P_y^q \, P_y^{\bar q} \right)
 \cos(\vphi_3 - \vphi_2) -
 \left( P_x^q \, P_y^{\bar q} - P_y^q \, P_x^{\bar q} \right)
 \sin(\vphi_3 - \vphi_2) \right] \Biggr\} \nonumber \>,
 \eea
where the amplitude ${\hat M}_1^0$ is zero because of helicity conservation and
 \bea
 |\hat M_{2}^{0}|^{2}&=&\frac{64}{27}\,g_s^4
 \left(\frac{\hat{u}}{\hat{t}}- \frac{9}{4} \,
 \frac{\hat{u}^2}{\hat{s}^2}\right) \nonumber \\
 |\hat M_{3}^{0}|^{2}&=&\frac{64}{27}\,g_s^4
 \left(\frac{\hat{t}}{\hat{u}}- \frac{9}{4}\,
 \frac{\hat{t}^2}{\hat{s}^2}\right) \label{mmqqgg} \\
 \hat M_{2}^{0} \hat M_{3}^{0}&=&\frac{64}{27}\,g_s^4
 \left(1-\frac{\hat{t}\hat{u}}{\hat{s}^2}\right)\,.\nonumber
 \eea
The expression for $\bar q q \to gg$ is obtained from (\ref{qqgg}) with the
replacements $q \leftrightarrow \bar q$ and $\vphi_2 \leftrightarrow \vphi_3$.

\vskip 6pt
 \noindent {\bf 7)} $g_a g_b \to q \bar q$ {\bf processes}
 \bea
 && \hskip -20pt \Sigma(S_A,S_B)^{g_a g_b \to q \bar q} =
 \frac 12 \, \hat D_{C/q}(z, k_{\perp C}) \>
 \hat f_{g/S_A}(x_a, \bfk_{\perp a}) \> \hat f_{g/S_B}(x_b, \bfk_{\perp b})
 \times \nonumber \\
 && \hskip 40pt \Biggl\{ \left( 1 - P_z^a \, P_z^b \right)
 \left( |\hat M^0_2|^2 + |\hat M^0_3|^2 \right) \label{ggqq} \\
 && \hskip 40pt + \> 2  \hat M^0_2 \, \hat M^0_3 \left[
 \left( \Ione^a \, \Ione^b + \Itwo^a \, \Itwo^b \right)
 \cos(\vphi_3 - \vphi_2) -
 \left( \Ione^a \, \Itwo^b - \Itwo^a \, \Ione^b \right)
 \sin(\vphi_3 - \vphi_2) \right] \Biggr\} \>, \nonumber
 \eea
and the relevant amplitudes are the same as in Eq.~(\ref{mmqqgg}), multiplied 
by the factor $9/64$. The expression for $gg \to \bar q q$ is obtained 
from Eq.~(\ref{ggqq}) with the replacements $q \leftrightarrow \bar q$ and 
$\vphi_2 \leftrightarrow \vphi_3$.

\vskip 6pt
\noindent {\bf 8)} $g_a g_b \to g_c g_d$ {\bf processes}
 \bea
 && \Sigma(S_A,S_B)^{g_a g_b \to g_c g_d} =
 \frac 12 \, \hat D_{C/g}(z, k_{\perp C}) \>
 \hat f_{g/S_A}(x_a, \bfk_{\perp a}) \> \hat f_{g/S_B}(x_b, \bfk_{\perp b})
 \times \nonumber \\
 && \quad\quad\quad\quad
 \Biggl\{ \left( |\hat M^0_1|^2 + |\hat M^0_2|^2 + |\hat M^0_3|^2 \right)
 + P_z^a \, P_z^b \left( |\hat M^0_1|^2 - |\hat M^0_2|^2 - |\hat M^0_3|^2
 \right) \nonumber \\
 && \quad\quad\quad\quad
 + \>  2 \hat M^0_2 \, \hat M^0_3 \left[ \left(
 \Ione^a \, \Ione^b + \Itwo^a \, \Itwo^b \right) \,
 \cos(\vphi_{3}-\vphi_{2}) + \left(
 \Itwo^a \, \Ione^b + \Ione^a \, \Itwo^b \right) \,
 \sin(\vphi_{3}-\vphi_{2})
 \right] \Biggr\} \nonumber \\
 && \quad\quad\quad\quad
 + \> \frac 12 \, \Delta^N \hat{D}_{C/\IoneG}(z, k_{\perp C}) \>
 \hat f_{g/S_A}(x_a, \bfk_{\perp a}) \> \hat f_{g/S_B}(x_b, \bfk_{\perp b}) 
 \times \label{gggg} \\
 && \quad\quad\quad\quad
 \Biggl\{ \hat M^0_1 \, \hat M^0_2 \, \left[
  \Ione^a \, \cos(\vphi_1 - \vphi_2 + 2\phi_C^H)
 + \Itwo^a \, \sin(\vphi_1 - \vphi_2 + 2\phi_C^H) \right] \nonumber \\
 && \quad\quad\quad\quad
 + \> \hat M^0_1 \, \hat M^0_3 \, \left[
  \Ione^b \, \cos(\vphi_1 - \vphi_3 + 2\phi_C^H)
 + \Itwo^b \, \sin(\vphi_1 - \vphi_3 + 2\phi_C^H) \right] \Biggr\} \>, 
\nonumber
 \eea
where
 \bea
 |\hat M_{1}^{0}|^{2}&=&\frac{9}{2}\,g_s^4\,\hat{s}^{2}
 \left(\frac{1}{\hat{t}^{2}}+\frac{1}{\hat{u}^{2}}+
 \frac{1}{\hat{t}\hat u}\right) \nonumber\\
 |\hat M_{2}^{0}|^{2}&=&\frac{9}{2}\,g_s^4\,
 \frac{\hat{u}^{2}}{\hat{s}^{2}}\left(1+\frac{\hat u}{\hat t}+
 \frac{\hat{u}^{2}}{\hat{t}^{2}}\right) \nonumber\\
 |\hat M_{3}^{0}|^{2}&=&\frac{9}{2}\,g_s^4\,
 \frac{\hat{t}^{2}}{\hat{s}^{2}}\left(1+\frac{\hat t}{\hat u}+
 \frac{\hat{t}^{2}}{\hat{u}^{2}}\right) \nonumber\\
 \hat M_{1}^{0} \hat M_{2}^{0}&=&\frac{9}{2}\,g_s^4
 \left(1+\frac{\hat u}{\hat t}+\frac{\hat{u}^{2}}{\hat{t}^{2}}\right)
 \label {mmgggg} \\
 \hat M_{1}^{0} \hat M_{3}^{0}&=&
 \frac{9}{2}\,g_s^4\left(1+\frac{\hat t}{\hat u}+
 \frac{\hat{t}^{2}}{\hat{u}^{2}}\right) \nonumber\\
 \hat M_{2}^{0} \hat M_{3}^{0}&=&
 \frac{9}{2}\,g_s^4\,\frac{1}{\hat s^{2}}\left(\hat{u}^{2}+\hat{t}^{2}+
 \hat{u}\hat{t}\right)\,.\nonumber
 \eea

\section{Polarized cross section and spin asymmetries}

Knowing the kernels $\Sigma(\SA,\SB)$, we could now proceed with the 
computation of any polarized cross section and spin asymmetry, according 
to our spin and $\bfk_\perp$ dependent factorization scheme,
\bea
\frac{E_C \, d\sigma^{(A,S_A) + (B,S_B) \to C + X}}
{d^{3} \bfp_C} &=& \sum_{a,b,c,d} \int \frac{dx_a \,
dx_b \, dz}{16 \pi^2 x_a x_b z^2  s} \;
d^2 \bfk_{\perp a} \, d^2 \bfk_{\perp b}\, d^3 \bfk_{\perp C}\,
\delta(\bm{k}_{\perp C} \cdot \hat{\bm{p}}_c) \, J(\bm{k}_{\perp C})
\nonumber \\
&\times& \Sigma(S_A,S_B)^{ab \to cd} 
(x_a, x_b, z, \bfk_{\perp a},\bfk_{\perp b}, \bfk_{\perp C})  
\> \delta(\hat s + \hat t + \hat u) \>, \label{gensig} 
\eea
where the sum over all kinds of partons leads to the 8 kernels 
$\Sigma(S_A,S_B)$ explicitely given in Eqs.~(\ref{qqqq})--(\ref{mmgggg}).

In the remaining of the paper we shall consider the unpolarized cross section 
and the transverse single spin asymmetry $A_N$ and show numerically how much 
different effects can contribute to their values. The single spin asymmetry 
$A_N$, measured in $\pup p \to \pi X$ scatterings, is defined as 
\be
A_N=\frac{d\sigma^\ua - d\sigma ^\da}{d\sigma^\ua + d\sigma ^\da}\,,
\ee
and requires the evaluation and integration of the quantities
\(\Sigma(\ua,0) - \Sigma(\da,0)\) in the numerator, and
\(\Sigma(\ua,0) + \Sigma(\da,0)\) in the denominator. Indeed, the difference
and sum of these kernels have to be evaluated for each elementary process
$ab \to cd$: we shall explicitly show the analytical formulae corresponding
to four channels only, which serve as examples; all the other contributions 
can be straightforwardly computed in a similar way.

For the numerator of the single spin asymmetry, for the process 
$\Aup \, B \to C \, X$, we consider explicitely the following channels:

\vskip 6pt
\noindent {\bf N-a)} $q_a q_b \to q_c q_d$
\bea
&&\hspace*{-1.6cm}[\Sigma(\ua,0) - \Sigma(\da,0)]^{q_a q_b \to q_c q_d} =
\nonumber \\
&\,& \frac{1}{2} \, \Delta \hf_{a/\Aup} (x_a,\bfk_{\perp a}) \,
\hf_{b/B}(x_b, k_{\perp b}) \,
\left[\,|{\hat M}_1^0|^2 + |{\hat M}_2^0|^2 + |{\hat M}_3^0|^2 \right] \,
\hat D _{C/c} (z, k_{\perp C}) \nonumber \\
&+&  2\,\left[ \Delta^- \hf^a_{s_y/\ua} (x_a, \bfk_{\perp a}) \, 
\cos(\varphi_3 -\varphi_2) -\Delta \hf^a_{s_x/\ua} (x_a,\bfk_{\perp a}) \, 
\sin(\varphi_3 -\varphi_2) \right] \,
\nonumber \\
&& \times \, \Delta \hf^b_{s_y/B}(x_b,\bfk_{\perp b})\,
{\hat M}_2^0 \, {\hat M}_3^0 \,\hat D _{C/c} (z, k_{\perp C})
\label{num-asym-qq} \\
&+& \left[ \Delta^- \hf^a_{s_y/\ua} (x_a, \bfk_{\perp a})\,
\cos(\varphi_1 -\varphi_2 + \phi_C^H)
- \Delta \hf^a_{s_x/\ua}(x_a,\bfk_{\perp a})\,
\sin(\varphi_1 -\varphi_2 + \phi_C^H) \right] \,\nonumber \\
&&\times \, \hf_{b/B}(x_b, k_{\perp b})\,
{\hat M}_1^0 \, {\hat M}_2^0 \, \Delta^N {\hat D}_{C/\cupar} (z, k_{\perp C})\,
\nonumber \\
&+&  \frac{1}{2} \, \Delta \hf_{a/\Aup} (x_a, \bfk_{\perp a})\,
\Delta \hf^b_{s_y/B} (x_b, \bfk_{\perp b}) \, 
\cos(\varphi_1 -\varphi_3 + \phi_C^H) \, {\hat M}_1^0 \, {\hat M}_3^0 \,
\Delta ^N {\hat D} _{C/\cupar} (z, k_{\perp C}) \nonumber 
\eea
\vskip 6pt
\noindent {\bf N-b)} $q \bar q \to g_c g_d$
\bea
&&\hspace*{-1.6cm}[\Sigma(\ua,0) - \Sigma(\da,0)]^{q \bar q \to g_c g_d} =
\nonumber \\
&\,& \frac{1}{2} \, \Delta \hf_{q/\Aup} (x_q,\bfk_{\perp q}) \,
\hf_{\bar q/B} (x_{\bar q}, k_{\perp \bar q}) \,
\left[ \, |{\hat M}_2^0|^2 + |{\hat M}_3^0|^2 \right] \,
\hat D _{C/g} (z, k_{\perp C}) \nonumber \\
&+&  2 \, \left[ \Delta^- \hf^q_{s_y/\ua} (x_q,\bfk_{\perp q}) \, 
\cos(\varphi_3 -\varphi_2)
-\Delta \hf^q_{s_x/\ua} (x_q, \bfk_{\perp q}) \, 
\sin(\varphi_3 -\varphi_2) \right] \,
\nonumber \\
&& \times \, \Delta \hf^{\bar q}_{s_y/B} (x_{\bar q}, \bfk_{\perp \bar q})\,
{\hat M}_2^0 \, {\hat M}_3^0 \,\hat D _{C/g} (z, k_{\perp C})
\label{num-asym-qbarq}
\eea
\vskip 6pt
\noindent {\bf N-c)} $qg \to qg$
\bea
&&\hspace*{-1.6cm}[\Sigma(\ua,0) - \Sigma(\da,0)]^{q g \to q g} =
\nonumber \\
&\,& \frac{1}{2} \, \Delta \hf_{q/\Aup} (x_q, \bfk_{\perp q}) \,
\hf_{g/B}(x_g, k_{\perp g}) \,
\left[ \, |{\hat M}_1^0|^2 + |{\hat M}_2^0|^2 \right] \,
\hat D _{C/q} (z, k_{\perp C}) \nonumber \\
&+&  \> \left[ \Delta^- \hf^q_{s_y/\ua} (x_q, \bfk_{\perp q}) \,
\cos(\varphi_1 -\varphi_2 +\phi_C^H)
- \Delta \hf^q_{s_x/\ua} (x_q,\bfk_{\perp q}) \,
\sin(\varphi_1 -\varphi_2 + \phi_C^H) \right] \,
\nonumber \\
&&\times \, \hf_{g/B}(x_g, k_{\perp g})\,
{\hat M}_1^0 \, {\hat M}_2^0 \, \Delta^N \hat{D}_{C/\qup}(z, k_{\perp C})
\label{num-asym-gq} 
\eea
\vskip 6pt
\noindent {\bf N-d)} $g_a g_b \to g_c g_d$
\bea
&&\hspace*{-1.4cm}[\Sigma(\ua,0) - \Sigma(\da,0)]^{g_ag_b\to g_cg_d} =
\nonumber \\
&\,& \frac{1}{2} \, \Delta \hf_{g/\Aup} (x_{a}, \bfk_{\perp a}) \,
\hf_{g/B}(x_{b}, k_{\perp b}) \,
\left[ \, |{\hat M}_1^0|^2 + |{\hat M}_2^0|^2 + |{\hat M}_3^0|^2 \right] \,
\hat D _{C/g} (z, k_{\perp C}) \nonumber \\
&+&  2 \, \left[ \Delta^- \hf^g_{\Ione/\ua} (x_{a}, \bfk_{\perp a}) \,
\cos(\varphi_3 -\varphi_2) +\Delta \hf^g_{\Itwo/\ua} (x_{a}, \bfk_{\perp a}) 
\, \sin(\varphi_3 -\varphi_2) \right] \, \nonumber \\
&& \times \, \Delta \hf^g_{\Ione/B}(x_b,\bfk_{\perp b})\,
{\hat M}_2^0 \, {\hat M}_3^0 \,\hat D _{C/g} (z, k_{\perp C})
\label{num-asym-gg} \\
&+& \> \left[ \Delta^- \hf^g_{\Ione/\ua} (x_{a}, \bfk_{\perp a})\,
\cos(\varphi_1 -\varphi_2 + 2 \phi_C^H)
 + \Delta \hf^g_{\Itwo/\ua} (x_{a}, \bfk_{\perp a})\,
\sin(\varphi_1 -\varphi_2 + 2\phi_C^H) \right] \nonumber \\
&& \times \, \hf_{g/B}(x_{b}, k_{\perp b})\,
{\hat M}_1^0 \, {\hat M}_2^0 \, \Delta^N \hat{D}_{C/\IoneG}(z, k_{\perp C})
\nonumber \\
&+& \frac 12 \, \Delta \hf_{g/\Aup} (x_{a},\bfk_{\perp a})\,
\Delta \hf^g_{\Ione/B}(x_{b}, \bfk_{\perp b})\,
\cos(\varphi_1 -\varphi_3 + 2\phi_C^H)\,
{\hat M}_1^0 \, {\hat M}_3^0 \,
\Delta^N \hat{D}_{C/\IoneG}(z, k_{\perp C}) \nonumber \,.
\eea

The above 4 cases have been obtained respectively from the kernels in
Eqs.~(\ref{qqqq}), (\ref{qqgg}), (\ref{gqgq}) and (\ref{gggg}), taking into 
account that
\bea
&& P_x^a \, \hf _{a/\ua} = - P_x^a \, \hf _{a/\da} \quad\quad\quad\>\,
\Itwo^g \hf_{g/\ua} = - \Itwo ^g \hf_{g/\da} \nonumber \\
&& P_y^a \, \hf _{a/\ua} - P_y^a \, \hf _{a/\da} = 2 \, \Delta^- \hf _{s_y/\ua}
\quad\quad\quad\>\,
\Ione^g \hf_{g/\ua} - \Ione^g \hf_{g/\da} =  2 \, \Delta^- \hf^g_{\Ione/\ua}
\nonumber \\
&& P_x^b \, \hf _{b/B} = \Delta \hf _{s_x/B} = 0 \quad\quad
   P_z^b \, \hf _{b/B} = \Delta \hf _{s_z/B} = 0 \quad\quad
   \Itwo ^g \hf_{g/B} = \Delta \hf _{\Itwo/B} = 0 \nonumber
\eea
as one can see from Eqs.~(\ref{A23}), (\ref{A46}), (\ref{Pxsa}),
(\ref{Pzsa}) and (\ref{imrho+-g}).

Let us inspect Eq.~(\ref{num-asym-qq}), which has an immediate partonic 
interpretation. The first line contains the Sivers effect, where the Sivers 
distribution function for quark $a$ appears in association with the 
unpolarized parton distribution function (PDF) for quark $b$, with the 
unpolarized elementary cross section and with the unpolarized fragmentation 
function (FF) for quark $c\,$;
the second and third lines correspond to the Boer-Mulders effect,
in which the Boer-Mulders PDF for quark $b$ is convoluted with
a complicated combination of distribution functions for quark $a$ which,
once integrated over the intrinsic transverse momentum $\bfk _{\perp a}$,
is somehow related to the transversity function $\Delta _T q (x_a)$ or
$h_1^a(x_a)$, and the unpolarized FF for quark $c\,$; the fourth and fifth 
lines contain the Collins term, coupled to transversity distributions, which 
was already extensively discussed in Ref.~\cite{noi} (notice that, exploiting 
Eqs.~(\ref{M-M0}), (\ref{phases}) and (\ref{A25}), one can explicitely 
show that this term exactly agrees with that in Eq. (56) of Ref.~\cite{noi}); 
finally the sixth line contains a ``mixed'' term in which all three effects 
(Sivers$\,\otimes\,$Boer-Mulders$\,\otimes\,$Collins) appear together.

Notice that Eq.~(\ref{num-asym-gg}), corresponding to $gg\to gg$ elementary 
scattering, has the same structure as Eq.~(\ref{num-asym-qq}), related to 
$qq \to qq$ elementary channel: while the Sivers function can be defined 
also for gluons (first line) the other terms correspond to linearly 
polarized gluons inside an unpolarized hadron (``Boer-Mulders-like''), 
to distributions of linearly polarized gluons inside a transversely polarized 
hadron (``transversity-like'') and to the fragmentation of linearly polarized 
gluons into an unpolarized hadron (``Collins-like'').  

In Eqs.~(\ref{num-asym-qbarq}) and (\ref{num-asym-gq}), related respectively 
to $q\bar q \to gg$ and $qg\to qg$ elementary scatterings, one can recognise 
the Sivers contribution (first line), the 
(transversity$\,\otimes\,$Boer-Mulders) 
[second and third lines of Eq.~(\ref{num-asym-qbarq})] and the  
(transversity$\,\otimes\,$Collins)
[second and third lines of Eq.~(\ref{num-asym-gq})] effects.

Concerning the denominator of the single spin asymmetry,
$d\sigma ^\ua + d\sigma ^\da = 2\, d\sigma ^{unp}$, the relevant quantity 
we have to calculate is $[\Sigma(\ua,0) + \Sigma(\da,0)]$ for each partonic 
process $ab\to cd$. We present explicit results for the same channels we 
have considered above.
\vskip 6pt
\noindent {\bf D-a)} $q_a q_b \to q_c q_d$
\bea
&&\hspace*{-1.8cm}[\Sigma(\ua,0) + \Sigma(\da,0)]^{q_aq_b\to q_cq_d} =
\nonumber \\
&\,& \hf_{a/A} (x_a, k_{\perp a}) \, \hf_{b/B} (x_b, k_{\perp b}) \,
\left[ \, |{\hat M}_1^0|^2 + |{\hat M}_2^0|^2 + |{\hat M}_3^0|^2 \right] \,
\hat D _{C/c} (z, k_{\perp C}) \nonumber \\
&+&  2\, \Delta \hf^a_{s_y/A} (x_a, \bfk_{\perp a}) \,
         \Delta \hf^b_{s_y/B} (x_b, \bfk_{\perp b})\, 
\cos(\varphi_3 -\varphi_2)\,
{\hat M}_2^0 \, {\hat M}_3^0 \,\hat D _{C/c} (z, k_{\perp C})
\label{den-asym-qq} \\
&+& \left[ \hf_{a/A} (x_a, k_{\perp a})\,
\Delta \hf^b_{s_y/B} (x_b,\bfk_{\perp b})\, 
\cos(\varphi_1 -\varphi_3 + \phi_C^H)\,
{\hat M}_1^0 \, {\hat M}_3^0 \nonumber \right. \\
&& \! + \, \left.
\Delta \hf^a_{s_y/A} (x_a,\bfk_{\perp a})\,
\hf_{b/B} (x_b, k_{\perp b})\, \cos(\varphi_1 -\varphi_2 + \phi_C^H)\,
{\hat M}_1^0 {\hat M}_2^0\, \right] \,
\Delta ^N {\hat D} _{C/\cupar} (z, k_{\perp C}) \nonumber 
\eea
\vskip 6pt
\noindent {\bf D-b)} $q \bar q \to g_c g_d$
\bea
&&\hspace*{-2.1cm}[\Sigma(\ua,0) + \Sigma(\da,0)]^{q\bar q \to g_c g_d} =
\nonumber \\
&\,& \hf_{q/A} (x_q, k_{\perp q}) \,
\hf_{\bar q/B} (x_{\bar q}, k_{\perp \bar q}) \,
\left[ \, |{\hat M}_2^0|^2 + |{\hat M}_3^0|^2 \right] \,
\hat D _{C/g} (z, k_{\perp C}) \label{den-asym-qbarq}\\
&+&  2\, \Delta \hf^q_{s_y/A} (x_q,\bfk_{\perp q}) \,
         \Delta \hf^{\bar q}_{s_y/B} (x_{\bar q},\bfk_{\perp \bar q})\, 
\cos(\varphi_3 -\varphi_2)\,
{\hat M}_2^0 \, {\hat M}_3^0 \,\hat D _{C/g} (z, k_{\perp C}) \nonumber
\eea
\vskip 6pt
\noindent {\bf D-c)} $qg \to qg$
\bea
&&\hspace*{-1.8cm}[\Sigma(\ua,0) + \Sigma(\da,0)]^{qg\to qg} =
\nonumber \\
&\,& \hf_{q/A} (x_q, k_{\perp q}) \, \hf_{g/B} (x_g, k_{\perp g}) \,
\left[ \, |{\hat M}_1^0|^2 + |{\hat M}_2^0|^2 \right] \,
\hat D _{C/q} (z, k_{\perp C}) \nonumber \\
&+&  \> \Delta \hf^q_{s_y/A} (x_q,\bfk_{\perp q}) \,
\hf_{g/B} (x_g, k_{\perp g})\, \cos(\varphi_1 -\varphi_2 + \phi_C^H)\,
{\hat M}_1^0 \, {\hat M}_2^0 \, \Delta^N \hat{D}_{C/\qup}(z, k_{\perp C})
\label{den-asym-gq}
\eea
\vskip 6pt
\noindent {\bf D-d)} $g_a g_b \to g_c g_d$
\bea
&&\hspace*{-1.8cm}[\Sigma(\ua,0) + \Sigma(\da,0)]^{g_ag_b\to g_cg_d} =
\nonumber \\
&\,& \hf_{g/A} (x_a, k_{\perp a}) \, \hf_{g/B}(x_b, k_{\perp b}) \,
\left[ \, |{\hat M}_1^0|^2 + |{\hat M}_2^0|^2 + |{\hat M}_3^0|^2 \right] \,
\hat D _{C/g} (z, k_{\perp C}) \nonumber \\
&+&  2\, \Delta \hf^a_{\Ione/A} (x_{a}, \bfk_{\perp a}) \,
         \Delta \hf^b_{\Ione/B} (x_{b}, \bfk_{\perp b})\,
         \cos(\varphi_3 -\varphi_2)\,
{\hat M}_2^0 \, {\hat M}_3^0 \,\hat D _{C/g} (z, k_{\perp C})
\label{den-asym-gg} \\
&+& \> \left[ \hf_{g/A} (x_{a}, k_{\perp a})\,
\Delta \hf^b_{\Ione/B}(x_{b}, \bfk_{\perp b})\,
\cos(\varphi_1 -\varphi_3 + 2\phi_C^H)\,
{\hat M}_1^0 \, {\hat M}_3^0 \right. \nonumber \\
&& \, + \, \left.
\Delta \hf^a_{\Ione/A}(x_{a},\bfk_{\perp a})\,
\hf_{g/B}(x_{b}, k_{\perp b})\, \cos(\varphi_1 -\varphi_2 + 2\,\phi_C^H)\,
{\hat M}_1^0 \, {\hat M}_2^0\, \right] \,
\Delta^N \hat{D}_{C/\IoneG}(z, k_{\perp C}) \nonumber \,.
\eea

As for the asymmetry numerator, Eqs.~(\ref{den-asym-qq}) and
(\ref{den-asym-gg}), corresponding respectively to the elementary 
scatterings $qq\to qq$ and $gg\to gg$, have the same overall structure. 
In this case, the first line corresponds to the usual unpolarized term, 
the second line to a double Boer-Mulders (or ``Boer-Mulders-like'') effect, 
whereas the third and fourth lines contain a mixed term in which the 
Boer-Mulders and Collins (or ``Boer-Mulders-like'' and ``Collins-like'') 
effects appear together. Regarding the elementary $q\bar q \to gg$ and 
$qg\to qg$ channels, Eqs.~(\ref{den-asym-qbarq}) and (\ref{den-asym-gq}) 
show that there are two terms contributing to the unpolarized cross section: 
the usual unpolarized term in both cases, the double Boer-Mulders effect 
for $q\bar q \to gg$ and the mixed 
(Boer-Mulders$\,\otimes\,$Collins) effect for $qg\to qg$.
 
It might be surprising to notice that several spin and $\bfk_\perp$ 
dependent mechanisms could also contribute to the unpolarized cross section.
Their numerical relevance will be studied in the next Section.   

\section{Numerical estimates of maximal contributions of single terms}

We have now explicit and comprehensive analytical formulae to
compute cross sections and spin asymmetries, coupling LO QCD interactions 
and soft physics; all this basic information is contained in the kernels,
as given in Sections III and IV, to be inserted into Eq.~(\ref{gensig}). 

These kernels, their differences and sums, contain many unknown functions
(the soft part), which we have interpreted in terms of parton polarizations 
and distribution or fragmentation amplitudes; we also give their 
expressions according to the notations of the Amsterdam group (Appendix C). 
In summary, there are, for each kind of partons, 8 different distribution 
functions and 2 fragmentation functions (into unpolarized hadrons). Out of 
these, only the unpolarized PDF, the helicity distributions and the 
unpolarized fragmentation functions (at least for pions) 
can be considered as rather well known, 
from experimental information gathered in inclusive and semi-inclusive Deep 
Inelastic Scattering processes. Some approximate information has been very 
recently extracted also on the quark Sivers distribution
\cite{noidis12, efr, werner} 
and Collins fragmentation functions~\cite{werner}.

This lack of information might induce the thought that any realistic 
evaluation of physical observables, through the scheme of Eq.~(\ref{gensig}),
is hopeless; nevertheless, such a scheme, in simplified versions,
has already been successfully used to compute transverse single spin 
asymmetries \cite{noi95,fu} and unpolarized cross sections \cite{fu}. 
Actually, its spin and $\bfk_{\perp}$ correlations are unique in order to 
understand and predict many observed and measurable spin effects.     

The way out of this worrying thought is naturally offered by the very structure
of our scheme and its exact kinematical formulation, and was already 
partially explored, concerning the contribution of the Collins mechanism,
in Ref.~\cite{noi}. The many phases appearing in the elementary interactions,
due to the noncollinear configurations [see Eq.~(\ref{M-M0})], once the 
integration over the non observable intrinsic motion is performed, lead to 
large cancellations of most contributions from the new unknown functions. 
One can realize that looking, for example, at Eq.~(\ref{num-asym-qq}) 
which shows the $qq \to qq$ contributions to the SSA $A_N$. Apart from the 
first line (the Sivers mechanism) all other terms contain the complicated 
$\varphi_1, \varphi_2$ and $\varphi_3$ phases, whose integration almost
completely cancel their numerical values.  
  
These effects can be shown in a quantitative way. We have evaluated the
different contributions to the SSA $A_N$ and to the unpolarized cross section, 
taking for each of the unknown functions their upper bounds, originating from 
basic principles (like $|P_j^a| \leq 1$). We have indeed largely overestimated 
each single contribution. More precisely, we have adopted the following 
strategy:
\begin{itemize}
\item 
We have followed Ref.~\cite{fu}, assuming a gaussian $k_\perp$ dependence 
for all distribution functions, with $\sqrt{\langle k_\perp^2 \rangle}$ 
= 0.8 GeV/$c$; 
the same $k_{\perp C}$ dependence as in Ref.~\cite{fu} has been assumed for 
the fragmentation functions. At relatively low $p_T$ the inclusion of
$\bm{k}_\perp$ effects 
might result in making one or more of the partonic Mandelstam 
variables smaller than a typical hadronic scale. In this case perturbation 
theory would break down. In order to avoid such a problem and extend our 
approach down to $p_T$ around 1-2 GeV$/c$, we have introduced a regulator 
mass, $\mu=0.8$ GeV, shifting all partonic Mandelstam variables, that is  
\be 
\hat t \to \hat t -\mu^2,\;\;\;\; \hat u \to \hat u
-\mu^2,\;\;\;\; \hat s \to \hat s + 2\mu^2\>.  
\ee
Concerning the potential ambiguity in the behaviour 
of the strong coupling constant, $\alpha_s(Q^2)$, in 
the low $Q^2$ regime, we adopt the prescription originally 
proposed by Shirkov and Solovtsov \cite{ss}. As renormalization and 
factorization scales $Q=\hat{p}^*_T/2$ is used, where $\hat{p}^*_T$ is the 
transverse momentum of the fragmenting parton in the partonic c.m. frame.
A comprehensive study of these and related aspects can be found in 
Ref.~\cite{fu}.

We only stress that, even if the magnitude of each contribution to the 
unpolarized cross sections, in particular at the smallest $p_T$ values,
is sensitive to these choices, their relative magnitudes (which are being 
studied here) are almost not affected at all. Moreover, for the SSA $A_N$ 
(which is a ratio of cross sections) this dependence is definitely strongly 
reduced. 
\item 
The unpolarized PDF have been taken 
from Ref.~\cite{pdf} and the fragmentation functions from Ref.~\cite{ff}. 
We have used Eq.~(\ref{gensig}), taking into account all its partonic 
contributions, and not only those shown as an example in Section IV. 
\item 
All unknown polarized distribution functions have been replaced 
with the corresponding unpolarized distributions. In some cases this is 
certainly an overestimate: for the transversity distribution it violates
the Soffer bound \cite{sof}.
\item
The Sivers and Collins functions have been chosen saturating their 
positivity bounds:
\be
\Delta^N \hf_{a/\Aup}(x_a, k_{\perp a}) = 2 \, \hf_{a/A}(x_a, k_{\perp a})    
\quad\quad
\Delta^N \hat D_{C/\qup}(z, k_{\perp C}) = 2 \, \hat D_{C/q}(z, k_{\perp C})
\,.
\ee
\item
Whenever different pieces could combine with different signs ({\it e.g.},
Sivers or Collins functions for different quark flavours), we have 
summed them {\it assuming the same sign}, in order to avoid any kind of 
cancellations not resulting from phase space integrations.   
\end{itemize}

Our results are shown in Figs. 1--5, and we shortly comment them.
%\begin{itemize}
%\item

Fig. 1 shows the different contributions to the unpolarized cross section,
see Eqs.~(\ref{den-asym-qq})--(\ref{den-asym-gg}) for guidance. 
Our numerical estimate is performed for $p \, p \to \pi^0 \, X$, in the 
kinematical region of the E704 experiment. Our result clearly proves that 
the usual contribution involving $f_{a/A} \otimes f_{b/B} \otimes D_{C/c}$ 
largely dominates; even assuming the polarized distributions as large as the 
unpolarized ones and summing additively all of them, their final contributions 
to the unpolarized cross section, after integration over all intrinsic 
$\bfk_\perp$, are at least one order of magnitude smaller. 
%\item

Fig. 2 shows again the different contributions to the unpolarized cross 
section, for $p \, p \to \pi^0 \, X$ processes, in the kinematical region 
of the STAR experiment at RHIC. The dominance of the usual 
$f_{a/A} \otimes f_{b/B} \otimes D_{C/c}$ term, in comparison with all
other contributions, is clear again; the second most important contribution,
the Boer-Mulders$\,\otimes\,$Collins term, is one order of magnitude smaller. 
%\item

In Fig. 3 we plot the different maximised contributions to $A_N$, for 
the E704 experimental configuration and $\pup p \to \pi^+ \, X$ processes,
for which very large values of $A_N$ have been measured \cite{e704}.
One sees that the Sivers mechanism is largely dominant, that some effects 
might originate from the Collins function and all other contributions are 
negligible. Notice that while the Sivers effect is maximised only in the 
choice of the Sivers function, the Collins contribution is maximised both in 
the choice of the Collins function and the transversity distribution. We have 
shown separately the quark and gluon Sivers contribution; there might be a
negative $x_F$ region where one could eventually gain some information on 
the (maximised) gluon Sivers function.        
%\item

In Fig. 4 we plot the different maximised contributions to $A_N$, for
the kinematical region of STAR-RHIC experiment, which also has measured 
non zero values of $A_N$ in $\pup p \to \pi^0 \, X$ processes \cite{star}. 
Again, the Sivers mechanism gives the largest contribution, some effects 
might remain from the Collins mechanism and all other contributions are 
negligible. At negative $x_F$ all contributions are vanishingly small. 
%\item

In Fig. 5 we plot the different maximised contributions to $A_N$, for 
the kinematical region of the proposed PAX experiment at GSI \cite{pax},
$\pup \bar p \to \pi^+ \, X$. The situation is similar to that for the
E704 case, with the difference that there might be, at large negative $x_F$,
a region where the (maximised) gluon Sivers function gives a sizeable 
contribution.  
%\end{itemize}
  
\begin{figure}
\includegraphics[angle=-90,width=10cm]{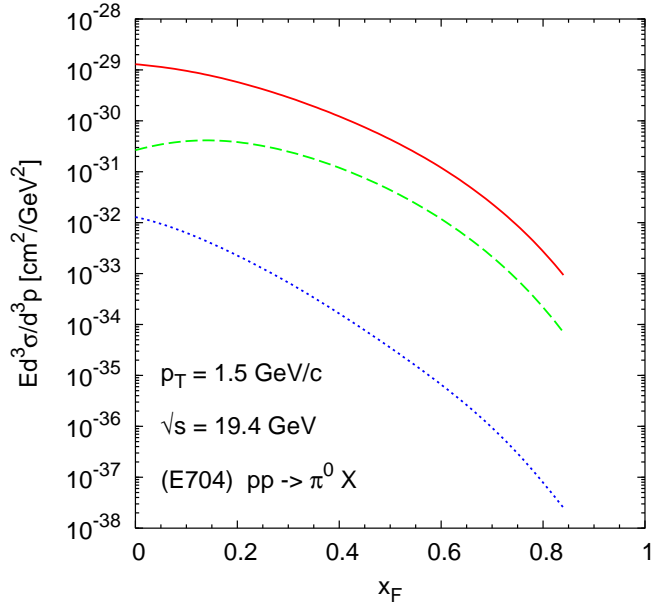}
\caption{\baselineskip=16pt
\label{unpe704} Different contributions to the unpolarized cross
section, plotted as a function of $x_F$, for $p \, p \to \pi^0 \, X$ processes
and E704 kinematics, as indicated in the plot. The three curves 
correspond to: {\it solid line} = usual unpolarized contribution; {\it dashed 
line} = Boer-Mulders $\otimes$ Collins; {\it dotted line} = Boer-Mulders 
$\otimes$ Boer-Mulders.}     
\end{figure}

\begin{figure}
\includegraphics[angle=-90,width=10cm]{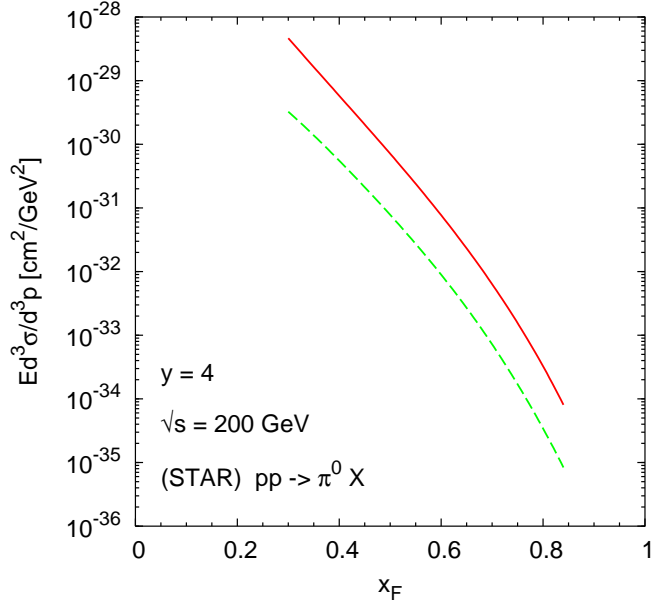}
\caption{\baselineskip=16pt
\label{unpstar} Different contributions to the unpolarized cross
section, plotted as a function of $x_F$, for $p \, p \to \pi^0 \, X$ processes
and STAR kinematics, as indicated in the plot. The 2 lines correspond 
to: {\it solid line} = usual unpolarized contribution; {\it dashed line} = 
Boer-Mulders $\otimes$ Collins. The Boer-Mulders $\otimes$ Boer-Mulders
contribution is not even noticeable at the scale of the figure.}       
\end{figure}

\begin{figure}
\includegraphics[angle=-90,width=10cm]{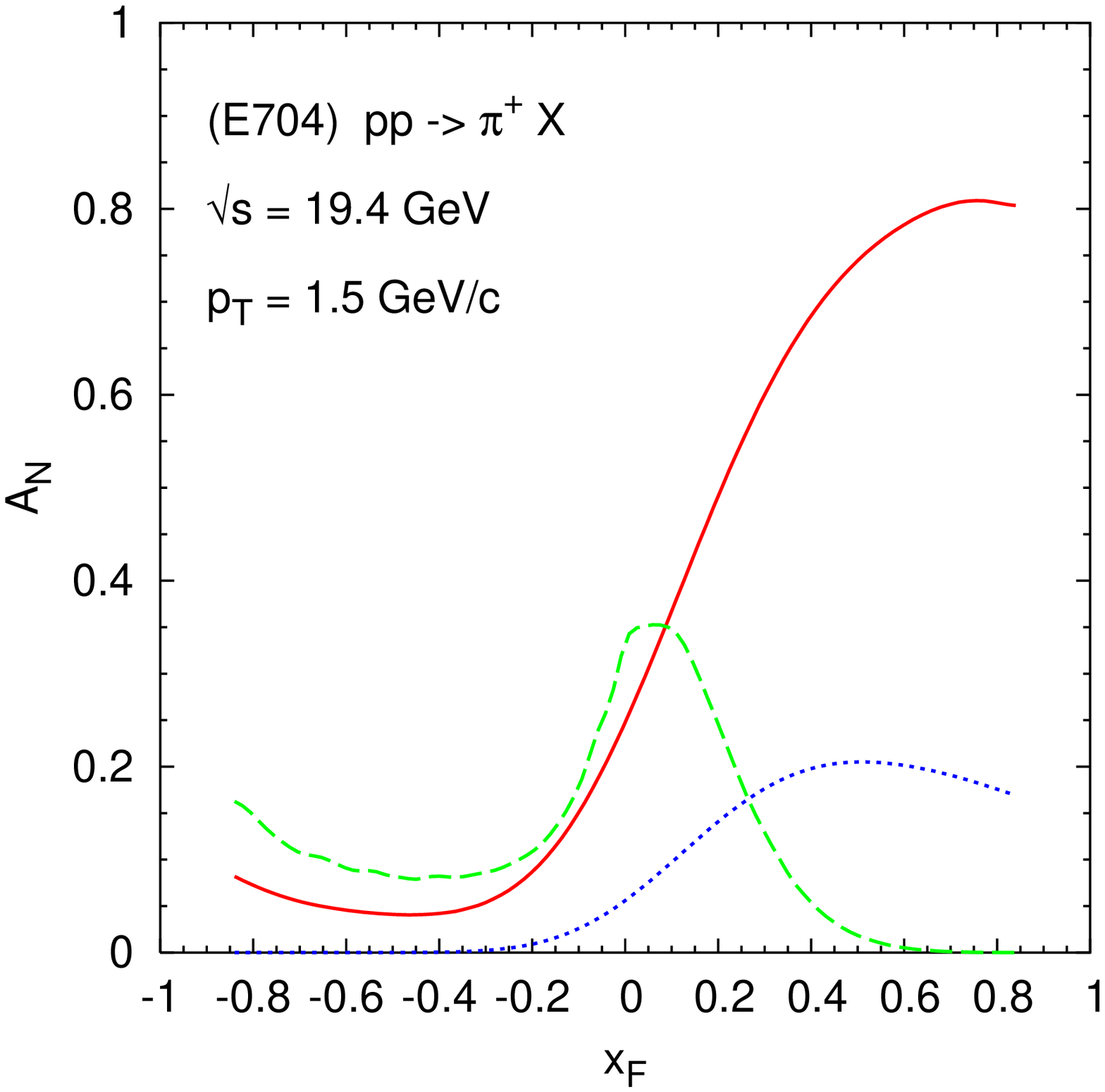}
\caption{\baselineskip=16pt
\label{asye704} Different contributions to $A_N$, plotted as a 
function of $x_F$, for $\pup p \to \pi^+ \, X$ processes and E704 kinematics.
The different lines correspond to: {\it solid line} = quark Sivers 
mechanism alone; {\it dashed line} = gluon Sivers mechanism alone; 
{\it dotted line} = transversity $\otimes$ Collins. All other contributions
are much smaller.} 
\end{figure}

\begin{figure}
\includegraphics[angle=-90,width=10cm]{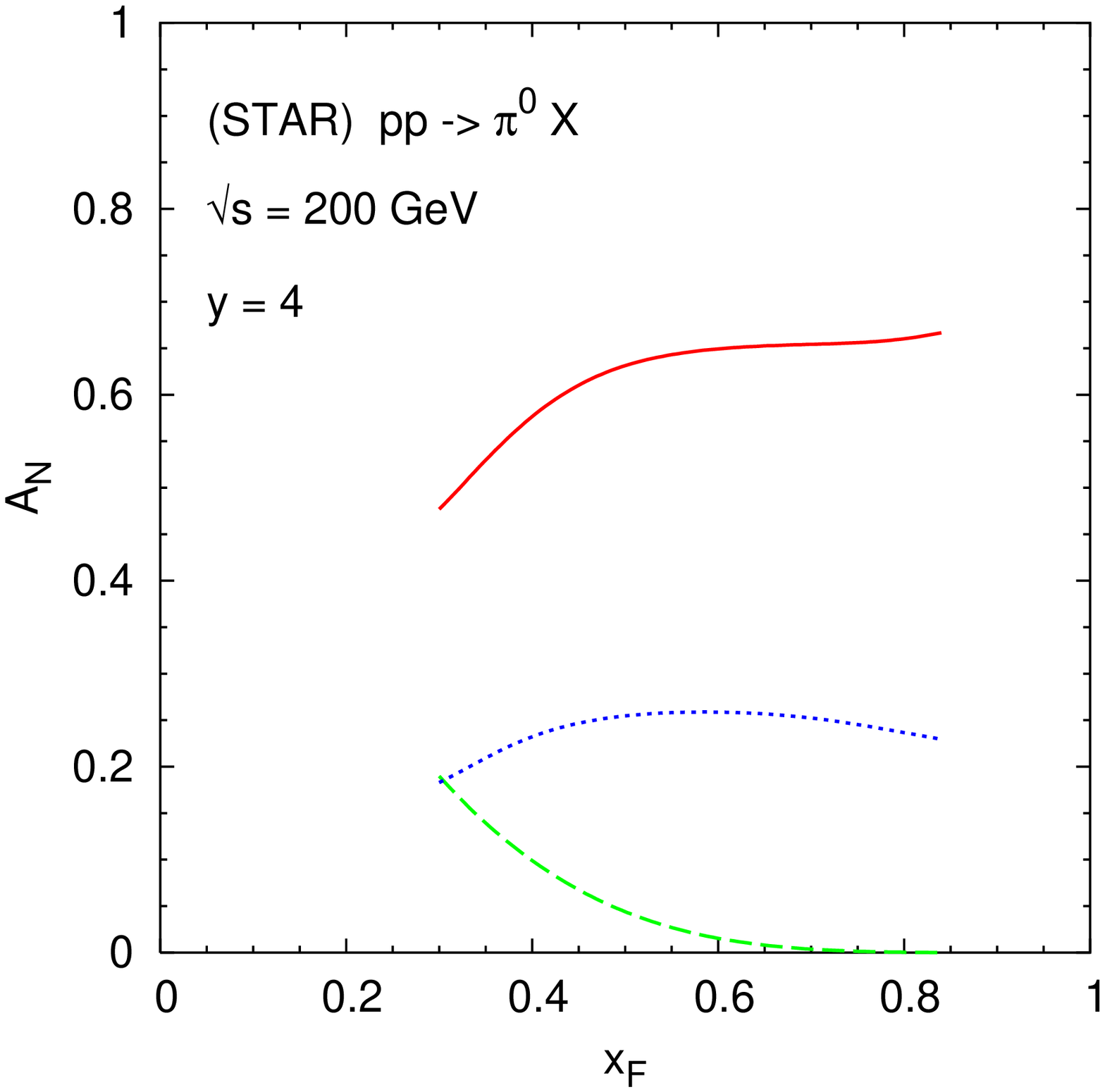}
\caption{\baselineskip=16pt
\label{asyestar} Different contributions to $A_N$, plotted as a 
function of $x_F$, for $\pup p \to \pi^0 \, X$ processes and STAR kinematics.
The different lines correspond to: {\it solid line} = quark Sivers 
mechanism alone; {\it dashed line} = gluon Sivers mechanism alone; 
{\it dotted line} = transversity $\otimes$ Collins. All other contributions
are much smaller.} 
\end{figure}

\begin{figure}
\includegraphics[angle=-90,width=10cm]{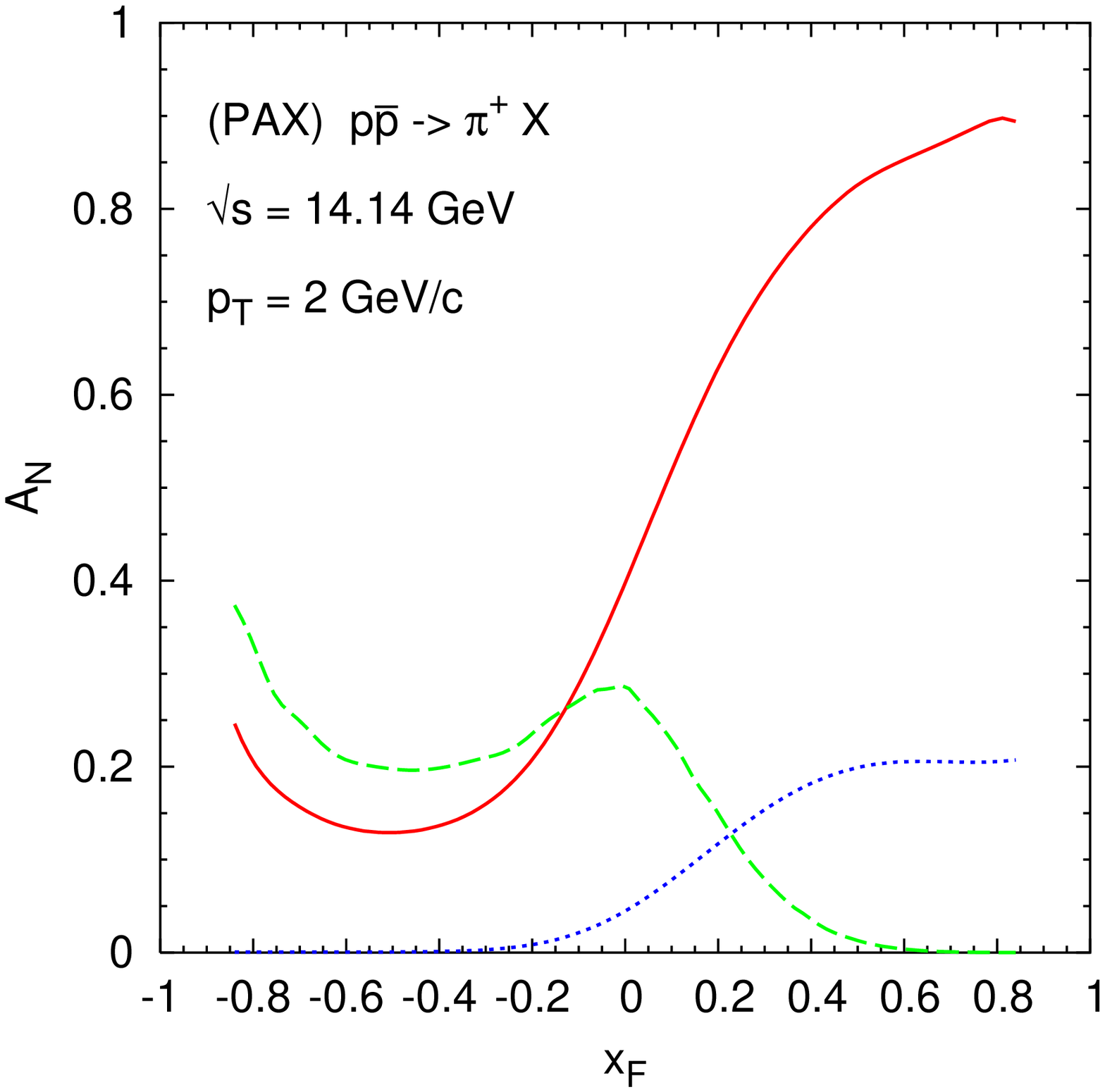}
\caption{\baselineskip=16pt
\label{asyepax} Different contributions to $A_N$, plotted as a 
function of $x_F$, for $\pup \bar p \to \pi^+ \, X$ processes and PAX 
kinematics, as indicated in the plot. The different lines correspond to: 
{\it solid line} = quark Sivers mechanism alone; {\it dashed line} = gluon 
Sivers mechanism alone; {\it dotted line} = transversity $\otimes$ Collins. 
All other contributions are much smaller.} 
\end{figure}

\section{Conclusions}

We have discussed in great detail a QCD based hard scattering formalism to 
compute unpolarized and polarized inclusive cross sections for the 
production of large $p_T$ particles in hadronic interactions. In the absence
of rigorous results, we have assumed a factorized 
scheme in which long distance non perturbative physics and short distance
pQCD interactions are separated and convoluted; such a factorization has been 
proven in collinear QCD, but has to be considered as a model when intrinsic
motion of partons -- effectively introducing higher-twist effects -- is 
allowed for. This is the first study in which the intrinsic $k_\perp$ of all
participating partons is taken into account. This intrinsic motion of partons, 
generated both by confinement 
and QCD dynamics, plays little or no role in unpolarized processes at very 
large energy, when all relevant momenta are much higher than the average 
$\langle k_\perp \rangle$; it is however crucial in unpolarized processes 
at intermediate energies \cite{fu} and, even more so, in the understanding 
of spin effects and polarized phenomena. For these, partonic 
spin--$\bfk_\perp$ correlations are of fundamental importance: an ever 
increasing number of spin experiments and spin measurements is proving that 
\cite{e704, star, herm}. 

Eq.~(\ref{gen1}) is our central point; it is essentially a QCD parton model,
in which LO (in $\alpha_s$) pQCD interactions couple to parton distribution 
and fragmentation functions; intrinsic motion is fully taken into account
in soft physics and in the elementary interactions. As it is well known,
this allows new soft partonic functions which would vanish in the collinear 
limit; however, it also introduces in the hard partonic interactions 
many $\bfk_\perp$-dependent phases, which strongly affect the convolution 
of the soft and hard parts. Luckily, it proves that such complicated
convolutions involving many phases and many soft functions, have the 
simplifying result of strongly suppressing most contributions to 
$(A,S_A) + (B,S_B) \to C + X$ inclusive processes. Concerning transverse
single spin asymmetries $A_N$, this leaves at work essentially only one 
spin--$\bfk_\perp$  correlation, namely the Sivers mechanism \cite{siv}. 
This allows to explain many measured and intriguing values of $A_N$ 
\cite{fu, noidis12}.  

We have fully discussed all soft functions, with attention to their
physical partonic interpretation, both in terms of polarized 
distribution and fragmentation functions and in terms of the amplitudes 
relating partonic and hadronic properties. We have also explicitely shown
the exact relationships between different notations widely used in the
literature; this should help in understanding and using the 
$\bfk_\perp$-dependent factorized scheme. Then, we have numerically shown 
the suppression of many contributions, both to the unpolarized cross section 
and the SSA $A_N$. This confirms and completes the work of Ref.~\cite{noi}.    
    
Many more applications of Eq.~(\ref{gen1}), modified to hold for different 
processes, can easily be foreseen. This has been done concerning the Sivers 
asymmetry in SIDIS processes \cite{noidis12} and can be extended to the SIDIS 
Collins asymmetry \cite{noiprep}; single and double spin asymmetries in single 
particle inclusive production and Drell-Yan processes can equally well be 
studied, and so on. Some information on Sivers and Collins functions is 
already available from ongoing experiments \cite{herm, belle} and more is
expected; a consistent understanding and computation of high energy spin 
effects, in the framework of a factorized QCD based model, is building up.

\acknowledgments

We acknowledge the support of the European Community--Research Infrastructure
Activity under the FP6 ``Structuring the European Research Area'' programme
(HadronPhysics, contract number RII3-CT-2004-506078). U.D., S.M. and F.M. 
acknowledge partial support by MIUR (Ministero dell'Istruzione, 
dell'Universit\`a e della Ricerca) under Cofinanziamento PRIN 2003. 
E.L. is grateful to 
the Royal Society of Edinburgh Auber Bequest for support and to 
INFN and the Sezione di Cagliari for periods of support 
and hospitality in the last years. M.A and U.D. are grateful to Imperial 
College for some hospitality.    

\appendix
\section{Detailed $\bfk_\perp$ kinematics}

We give here, for completeness and the reader's convenience, a detailed 
account of the partonic kinematics with the full inclusion of all transverse 
momenta, following Refs.~\cite{cont} and \cite{fu}. As throughout the paper, 
we consider the hadronic reaction $A \, B \to C \, X$ in the $AB$ center of 
mass frame with $A$ moving along the positive $Z_{cm}$-axis and we fix the 
scattering plane as the $(XZ)_{cm}$ plane. We neglect all masses, both the
hadronic and the partonic ones. 

The 4-momenta of hadrons $A,B,C$ then read
\be 
p_A^\mu = \frac{\sqrt s}{2}\,(1,\,0,\,0,\,1) \quad\quad 
p_B^\mu = \frac{\sqrt s}{2}\,(1,\,0,\,0,\,-1) \quad\quad 
p_C^\mu = (E_C,\,p_T,\,0,\,p_L) \,,
\ee 
with $E_C = \sqrt{p_T^2 + p_L^2}$ and $s=(p_A + p_B)^2$.

For massless partons $a,b$ inside hadrons $A,B$ we introduce light-cone 
momentum fractions $x_a = p_a^+/p_A^+ = (p_a^0 + p_a^3)/(p_A^0 + p_A^3)$, 
$x_b = p_b^+/p_B^+ = (p_b^0 - p_b^3)/(p_B^0 - p_B^3)$ and the transverse 
momenta $\bfk_{\perp a}$, $\bfk_{\perp b}$. Their four-momenta then read 
\bea
\label{papb}
p_a^\mu & = & x_a \, \frac{\sqrt s}{2} 
\left( 1 + \frac{k_{\perp a}^2}{x_a^2 \, s}, \> 
\frac{2k_{\perp a}}{x_a \sqrt s} \, \cos\phi_a, \> 
\frac{2k_{\perp a}}{x_a \sqrt s} \, \sin\phi_a, \> 
1 - \frac{k_{\perp a}^2}{x_a^2 \, s} \right) \nonumber\\
p_b^\mu & = & x_b \, \frac{\sqrt s}{2} 
\left( 1 + \frac{k_{\perp b}^2}{x_b^2 \, s}, \> 
\frac{2k_{\perp b}}{x_b \sqrt s} \, \cos\phi_b, \> 
\frac{2k_{\perp b}}{x_b \sqrt s} \, \sin\phi_b, \> 
- 1 + \frac{k_{\perp b}^2}{x_b^2 \, s} \right) \>, 
\eea 
where $k_{\perp a,b} = |\bfk_{\perp a,b}|$  and $\phi_{a,b}$
are the azimuthal angles of parton $a,b$ three-momenta in the hadronic
c.m. frame.

The four-momentum of the fragmenting parton $c$ is given in terms of the
observed hadron momentum $p_C^\mu$, of the light-cone momentum fraction 
$z=p_C^+/p_c^+$ and of the transverse momentum $\bfk_{\perp C}$ of hadron 
$C$ with respect to parton $c$ light-cone direction. In the hadronic 
c.m.~frame, we write in general $\bfk_{\perp C}$ as: 
\be \label{kc} 
\bfk_{\perp C} = k_{\perp C} \,
(\sin\theta_{k_{\perp C}}\cos\phi_{k_{\perp C}}, \>
 \sin\theta_{k_{\perp C}}\sin\phi_{k_{\perp C}}, \> 
 \cos\theta_{k_{\perp C}}) \,, 
\ee 
and impose the orthogonality condition $\bfk_{\perp C} \cdot \bfp_c = 0$
via the $\delta$-function $\delta(\bfk_{\perp C} \cdot \hat{\bfp}_c)$, where 
$\hat{\bfp}_c$ is the unit vector along the direction of motion of parton $c$. 
The parton four-momentum, $p_c^\mu = (E_c,\bfp_c)$, can then be written as 
\bea
\bfp_c 
& = &
\frac{E_c}{\sqrt{E_C^2 - k_{\perp C}^2}} \, 
(p_T - k_{\perp C} \sin\theta_{k_{\perp C}} \cos\phi_{k_{\perp C}}, \>
- k_{\perp C} \sin\theta_{k_{\perp C}} \sin\phi_{k_{\perp C}}, \> 
p_L - k_{\perp C} \cos\theta_{k_{\perp C}} )  \nonumber \\
\label{Ec}
E_c & = & \frac{E_C + \sqrt{E_C^2 - k_{\perp C}^2}}{2z} \>,  
\eea
and the orthogonality condition $\bfk_{\perp C} \cdot \bfp_c=0$ implies 
\bea
\label{dephik}
d^3 \bfk_{\perp C} \, \delta(\bfk_{\perp C} \cdot \hat{\bfp}_c) & = & 
k_{\perp C} \, dk_{\perp C} \, d\theta_{k_{\perp C}} \, d\phi_{k_{\perp C}} \,
\frac{\sqrt{E_C^2 - k_{\perp C}^2}}{p_T \, \sin\phi^0_{k_{\perp C}}} 
\nonumber \\
&\times& \label{2d}
\left[ \delta(\phi_{k_{\perp C}} - \phi^0_{k_{\perp C}}) + 
\delta(\phi_{k_{\perp C}} - (2\pi - \phi^0_{k_{\perp C}}) ) \right] \,,\\
\label{phik}
\cos\phi^0_{k_{\perp C}} & = &
\frac{k_{\perp C} - p_L \cos\theta_{k_{\perp C}}}
     {p_T \sin\theta_{k_{\perp C}}}, \;\; 0 \le \phi^0_{k_{\perp C}} 
\le \pi \>.  
\eea
This allows to perform directly the integration over $\phi_{k_{\perp C}}$
(notice that there are two possible solutions to be considered).

The factor $J(z, k_{\perp C})$ entering our basic factorization formula, 
Eq.~(\ref{gen1}), is the Jacobian factor connecting the parton $c$ to
hadron $C$ invariant phase space, defined as 
\be
\frac{d^3\bm{p}_c}{E_c} =
\frac{1}{z^2}\, J(z, k_{\perp C})\frac{d^3\bm{p}_C}{E_C} \,, 
\ee
which for collinear and massless particles reduces simply to $J = 1$.  
With intrinsic motion, for massless partons and hadrons:
\be
\label{jac}
J(z, k_{\perp C}) \equiv J(k_{\perp C}) = 
\frac{ \left(E_C + \sqrt{E_C^2 - k_{\perp C}^2} \right)^2}
{4 \, (E_C^2 - k_{\perp C}^2)} \, \cdot
\ee

With the expression of parton momenta given in Eqs.~(\ref{papb}) and
(\ref{Ec}) one can calculate the partonic Mandelstam invariants:
\bea
\label{shat}
\hat s = (p_a+p_b)^2 & = & x_a x_b s \left[1 - 2 \frac{k_{\perp a}
    k_{\perp b}}{x_ax_b s} \cos(\phi_a-\phi_b) + 
\frac{k_{\perp a}^2 k_{\perp b}^2}{x_a^2 x_b^2 s^2}
\right] \,,\\ 
\hat t = (p_a - p_c)^2 & = & \frac{T}{z} \,,\\ 
\hat u = (p_b - p_c)^2 & = & \frac{U}{z} \,,\\
\label{xc}
\hat s\, \delta(\hat s + \hat t + \hat u) & = & z\, \delta\left(z
+ \frac{T+U}{\hat s}\right) \,, 
\eea 
where 
\bea
T & = & - x_a \, \sqrt {s} \, 
\frac{E_C + \sqrt{E_C^2 - k_{\perp C}^2}}{2\sqrt{E_C^2 - k_{\perp C}^2}} 
\nonumber \\ 
& \times & \Bigg\{ \left( 1 + \frac{k_{\perp a}^2}{x_a^2 \, s} \right)
\sqrt{E_C^2 - k_{\perp C}^2} 
- \frac{2 k_{\perp a}}{x_{a} \sqrt s} \>
\cos\phi_a (p_T - k_{\perp C} \sin\theta_{k_{\perp C}} \cos\phi_{k_{\perp C}}) 
\nonumber \\
& + & \frac{2 k_{\perp a}}{x_{a} \sqrt  s} \>
k_{\perp C} \, \sin\phi_a \sin\theta_{k_{\perp C}} \sin\phi_{k_{\perp C}} 
- \left( 1 - \frac{k_{\perp a}^2}{x_a^2 \, s} \right) 
(p_L - k_{\perp C} \cos\theta_{k_{\perp C}}) \Bigg\} \\
U & = & - x_b \, \sqrt {s} \,
\frac{E_C + \sqrt{E_C^2 - k_{\perp C}^2}}{2\sqrt{E_C^2 - k_{\perp C}^2}} 
\nonumber \\ 
& \times & \Bigg\{ \left( 1 + \frac{k_{\perp b}^2}{x_b^2 \, s} \right)
\sqrt{E_C^2 -k_{\perp C}^2}  
- \frac{2 k_{\perp b}}{x_{b} \sqrt s} \>
\cos\phi_b (p_T - k_{\perp C} \sin\theta_{k_{\perp C}} \cos\phi_{k_{\perp C}}) 
\nonumber \\
& + & \frac{2 k_{\perp b}}{x_{b} \sqrt s} \>
k_{\perp C} \, \sin\phi_b \sin\theta_{k_{\perp C}} \sin\phi_{k_{\perp C}} 
+ \left( 1 - \frac{k_{\perp b}^2}{x_b^2 \, s} \right) 
(p_L - k_{\perp C} \cos\theta_{k_{\perp C}}) \Bigg\} \>.
\eea

The phase space integrations must obey some constraints, originating 
from physical requests. Besides the trivial bounds 
$0 < x_{a,b}, z < 1$, $0 \le \phi_{a,b} \le 2\pi$ and 
$0 \le \theta_{k_{\perp C}} \le \pi$, we require that, even including 
intrinsic transverse momentum effects, $a)$ each parton keeps moving along the
same direction as its parent hadron, $ \bfp_{a(b)} \cdot \bfP_{A(B)} >
0$, and $b)$ the parton energy is not larger than the parent hadron
energy, $E_{a(b)} \le E_{A(B)}$.  This implies the following bounds
\be
\label{kakblim}
k_{\perp a(b)}/\sqrt s < \min \left[x_{a(b)}, \> \sqrt{x_{a(b)}(1-x_{a(b)})}
\right]\>.  
\ee 

Analogously, for the fragmentation process $c\to C+X$
we require $\bfp_c\cdot \bfP_C> 0$ and $E_C \le E_c$ (both fulfilled
by Eq.~(\ref{Ec}), where we have consistently disregarded the solution
$E_c = \left[ E_C - \sqrt{E_C^2 - k_{\perp C}^2}\right]/(2z)$).  
The last constraint implies the following bound on $k_{\perp C}$, at fixed 
$z$ values: 
\be
\label{kclim}
k_{\perp C}/ E_C \le 1 \;\, (z \le 1/2) \quad\quad k_{\perp C}/E_C \le
2\sqrt{z(1-z)} \;\, (z > 1/2)\>.  
\ee
By requiring $|\cos\phi^0_{k_{\perp C}}|\leq 1$, see Eq.~(\ref{phik}), we 
have a further constraint on $k_{\perp C}$, at fixed $\theta_{k_{\perp C}}$, 
namely 
\be
\label{kclim2}
p_L\cos\theta_{k_{\perp C}} - p_T \sin\theta_{k_{\perp C}} \le 
k_{\perp C} \le p_L \cos\theta_{k_{\perp C}} + p_T \sin\theta_{k_{\perp C}}
\>.  
\ee

The partonic helicity amplitudes are computed according to 
Eqs.~(\ref{M-M0})--(\ref{phases}); the explicit expressions, in terms of 
the Mandelstam variables, of the relevant combinations of the $\hat M^0$  
amplitudes are given in Section III of the text. The phases $\varphi_i$ 
are defined in Eq.~(\ref{phases}). For processes involving only quarks 
and antiquarks they read:
\bea
\varphi_1 &=& -\frac 12 \left( \xi_a + \xi _b - \xi _c - \xi _d \right) 
\nonumber \\
\varphi_2 &=& \frac 12 \left( \xi_a - \xi _b - \xi _c + \xi _d \right)
           + \tilde \xi_a - \tilde \xi_c - \phi^{\prime\prime}_c
\label{phasesa} \\
\varphi_3 &=& \frac 12 \left( - \xi_a + \xi _b - \xi _c + \xi _d \right)
           - \tilde \xi_a - \tilde \xi_c + \phi^{\prime\prime}_c \>.
\nonumber
\eea
Similarly for processes involving also gluons.  
 
All terms appearing in the above phases are discussed and can be found 
in Ref.~\cite{noi}; we report them here for convenience and self-consistency 
of the paper:
\bea
\cos \xi _j &=&
\frac{\cos \theta _q \sin \theta _j - \sin \theta _q \cos \theta _j
      \cos(\phi _q - \phi _j)}{\sin \theta _{q p_j}} \label{cosxi} \\
\sin \xi _j &=&
\frac{\sin \theta _q \sin(\phi _q - \phi _j) }{\sin \theta _{q p_j}} \> \cdot
\label{sinxi}
\eea
All angles refer to the overall $AB$ c.m. frame. $\theta_j$ and $\phi_j$ 
($j =a,b,c,d$) are respectively the polar and azimuthal angles of the partons, 
while $\theta_q$ and $\phi_q$ are the polar and azimuthal angle of the 
vector $\bfq = \bfp_a + \bfp_b$. Here and in the next equations  
$\theta_{q p_j}$ ($0 \leq \theta _{q p_j} \leq \pi$) denotes in general
the angle between the two vectors $\bfq$ and $\bfp_j$.

The $\tilde \xi_j$ angles ($j =a,b,c,d$) are given by
\be
\tilde \xi _j = \eta ^\prime _j + \xi ^\prime _j \> ,
\ee
where
\bea
\cos \xi ^\prime _j &=&
\frac{ \cos \theta _q \sin \theta ^\prime _j -
       \sin \theta  _q \cos \theta ^\prime _j
       \cos(\phi _q - \phi ^\prime _j)}
     { \sin \theta _{q p^\prime _j}} \label{cosxip} \\
\sin \xi ^\prime _j &=&
\frac{ - \sin \theta _q
       \sin(\phi _q - \phi ^\prime _j)}
     { \sin \theta _{q p^\prime _j}} \label{denxip} \>;
\eea
\bea
\cos \eta ^\prime _j &=&
\frac{\cos \theta ^\prime _a  - \cos \theta ^\prime _j
      \cos \theta _{p^\prime_a p^\prime _j} }
     {\sin \theta ^\prime _j \sin \theta _{p^\prime _a p^\prime _j}}
\label{cosetap} \\
\sin \eta ^\prime _j &=&
\frac{\sin \theta ^\prime _a \sin(\phi ^\prime _a - \phi ^\prime _j)}
     {\sin \theta _{p^\prime _a p^\prime _j}}
\label{sinetap} \> \cdot
\eea
The primed angles $(\theta^\prime_j , \phi^\prime_j)$ are obtained via
\be
\bfp_i^{\prime} = \bfp_i - \frac{\bm{q}}{q^0 + \sqrt{q^2}}
\left( \frac {p_i \cdot q}{\sqrt{q^2}} + p_i^0 \right) \label{SS'}
\ee
where $i = a,b,c,d$ and $q^\mu = (q^0, \bm{q}) = p_a^\mu + p_b^\mu$.

The last angle appearing in Eqs.~(\ref{phasesa}) is $\phi^{\prime\prime}_c$,
given by:
\be
\tan \phi^{\prime\prime}_c =
\frac{\sin \theta ^\prime _c \sin(\phi ^\prime _c - \phi^\prime_a) }
     {\sin \theta ^\prime _c \cos(\phi ^\prime _c - \phi^\prime_a)
      \cos \theta ^\prime _a -
      \cos \theta ^\prime _c \sin \theta ^\prime _a}\,\cdot
\label{phi''}
\ee

Finally, the angle $\phi^H_C$ appearing in the fragmentation amplitudes,
Eq.~(\ref{fragphi}), is given, in terms of our integration and overall 
variables, by:
\be 
\tan \phi^H_C = \mp \frac{p_T}{\sqrt{E^2_C - k^2_{\perp C}}} \, 
\sqrt{1 - \left( \frac{k_{\perp C} - p_L \cos \theta_{k_{\perp C}}} 
{p_T \sin \theta _{k_{\perp C}}} \right)^2} \,
\tan \theta _{k_{\perp C}}\,, \label{phiC} 
\ee
where the $\mp$ signs refer, respectively, to the first and second 
$\delta$-function terms in Eq.~(\ref{2d}). 

\section{Parton polarizations and distribution amplitudes}

An alternative simple physical interpretation can be given to the distribution 
functions $(P^a_j \hat f_{a/A,S_A}) = \Delta \hat f ^a_{s_j/S_A}$
by making use of the helicity amplitudes
${\hat{\cal F}}_{\la^{\,}_a, \la^{\,}_{X_A};
\la^{\,}_A}$, which describe the soft process $A \to a + X$. This is the 
approach used in Refs.~\cite{noi,noi95}. Since the partonic distribution 
is usually regarded, at LO, as the inclusive cross section for this process, 
the helicity density matrix of parton $a$ inside hadron $A$ with spin $S_A$
and polarization vector $\bfP^A$ can be written as
\bea
\rho_{\la^{\,}_a, \la^{\prime}_a}^{a/A,S_A} \>
\hat f_{a/A,S_A}(x_a,\bfk_{\perp a})
&=& \sum _{\la^{\,}_A, \la^{\prime}_A}
\rho_{\la^{\,}_A, \la^{\prime}_A}^{A,S_A}
\sumint_{X_A, \la_{X_A}} \!\!\!\!\!
{\hat{\cal F}}_{\la^{\,}_a, \la^{\,}_{X_A};
\la^{\,}_A} \, {\hat{\cal F}}^*_{\la^{\prime}_a,\la^{\,}_{X_A}; \la^{\prime}_A}
\nonumber \\
&\equiv& \sum _{\la^{\,}_A, \la^{\prime}_A}
\rho_{\la^{\,}_A, \la^{\prime}_A}^{A,S_A} \>
\hat F_{\la^{\,}_A, \la^{\prime}_A}^{\la^{\,}_a,\la^{\prime}_a} \>,\label{defF}
\eea
having defined
\be
\hat{F}_{\la^{\,}_A, \la^{\prime}_A}^{\la^{\,}_a,\la^{\prime}_a} \equiv \>
\sumint_{X_A, \la_{X_A}} \!\!\!\!\!\!
{\hat{\cal F}}_{\la^{\,}_a,\la^{\,}_{X_A};\la^{\,}_A} \,
{\hat{\cal F}}^*_{\la^{\prime}_a,\la^{\,}_{X_A}; \la^{\prime}_A} \>,
\label{defFF}
\ee
where the $\sumint_{X_A, \la_{X_A}}\!\!\!$ stands for a spin sum and phase
space integration over all undetected remnants of hadron $A$, considered
as a system $X_A$, and the $\hat{\cal F}$'s are the {\it helicity distribution
amplitudes} for the $A \to a + X$ process.

Eq.~(\ref{defF}) relates the helicity density matrix of parton
$a$, see Eq.~(\ref{rho-a}), to the helicity density matrix of hadron $A$,
given by
\be
\rho_{\la^{\,}_A, \la^{\prime}_A}^{A,S_A} =
\frac{1}{2}\,{\left(
\begin{array}{cc}
1+P^A_Z & P^A_X - i P^A_Y \\
 P^A_X + i P^A_Y & 1-P^A_Z
\end{array}
\right)}_{\!\!\!A,S_A}=
\frac{1}{2}\,{\left(
\begin{array}{cc}
1+P^A_L & P^A_T \, e^{-i\phi_{S_A}}  \\
P^A_T \, e^{i\phi_{S_A}} & 1-P^A_L
\end{array}
\right)}_{\!\!\!A,S_A}
\,,
\label{rho-A}
\ee
where $\bfP ^A = (P^A_T \cos \phi_{S_A}, P^A_T \sin \phi_{S_A},  P^A_L)$ is
hadron $A$ polarization vector and  $\phi_{S_A}$ its azimuthal angle, defined 
in the helicity reference frame of hadron $A$. Notice that, in this Appendix,
we consider the most general case in which the transverse polarization of
hadron $A$ can be along any direction $\phi_{S_A}$ in the $XY$ plane, whereas 
in Section \ref{quark-sect} and throughout the paper the specific choice was 
made of fixing the transverse polarization of hadron $A$ along the $Y$-axis, 
{\it i.e.} $\ua\; = S_Y$, which corresponds to $\phi_{S_A}=\pi/2$.

The distribution amplitudes $\hat{\cal F}$ depend on the parton light-cone 
momentum fraction $x_a$ and on its intrinsic transverse momentum 
$\bfk_{\perp a}$, with modulus $k_{\perp a}$ and azimuthal angle $\phi_a$:
\be
\hat{\cal F}_{\la^{\,}_a,\la^{\,}_{X_A}; \la^{\,}_A}(x_a, \bfk_{\perp a})
= {\cal F}_{\la^{\,}_a,\la^{\,}_{X_A}; \la^{\,}_A}(x_a, k_{\perp a}) \>
{\rm exp}[i\la^{\,}_A \phi_a] \>,
\label{dampphi}
\ee
so that 
\be
\hat F_{\la^{\,}_A,\la^{\prime}_A}^{\la^{\,}_a,\la^{\prime}_a}(x_a,
\bfk_{\perp a})
= F_{\la^{\,}_A,\la^{\prime}_A}^{\la^{\,}_a,\la^{\prime}_a}(x_a, k_{\perp a})
\> {\rm exp}[i(\la^{\,}_A - \la^{\prime}_A)\phi_a] \>. \label{fft-ff}
\ee
$F_{\la^{\,}_A,\la^{\prime}_A}^{\la^{\,}_a,\la^{\prime}_a}(x_a, k_{\perp a})$
has the same definition as
$\hat{F}_{\la^{\,}_A,\la^{\prime}_A}^{\la^{\,}_a,\la^{\prime}_a}
(x_a, \bfk_{\perp a})$, Eq.~(\ref{defFF}), with $\hat{\cal F}$ replaced by
${\cal F}$, and does not depend on phases anymore.

The parity properties of
${\cal F}_{\la^{\,}_a, \la^{\,}_{X_A}; \la^{\,}_A}(x_a, k_{\perp a})$ are the
usual ones valid for helicity amplitudes in the $\phi_a=0$ plane \cite{elliot},
\be
{\cal F}_{-\la^{\,}_a,-\la^{\,}_{X_A}; -\la^{\,}_A} =
\eta \, (-1)^{S_A - s_a - S_{X_A}} \> (-1)^{\la^{\,}_A - \la^{\,}_a +
\la^{\,}_{X_A}}
\> {\cal F}_{\la^{\,}_a,\la^{\,}_{X_A}; \la^{\,}_A} \>, \label{parF}
\ee
where $\eta$ is an intrinsic parity factor such that $\eta^2 = 1$.
These imply:
\be
{F}_{-\la^{\,}_{A},-\la^{\prime}_A}^{ -\la^{\,}_a,-\la^{\prime}_{a}}
= (-1)^{2(S_A -s_a)} \>
(-1)^{(\la^{\,}_A -\la^{\,}_a) + (\la^{\prime}_A -\la^{\prime}_a)} \>
{F}_{\la^{\,}_{A},\la^{\prime}_A}^{ \la^{\,}_a,\la^{\prime}_{a}}
\,.
\label{parFF}
\ee
Notice that, for $S_A=1/2$, the factor $(-1)^{2(S_A -s_a)}$ is positive if 
parton $a$ is a quark and negative if it is a gluon; consequently, some parity 
relations are different according to the nature of the parton involved. For 
this reason we shall treat quark and gluon distribution functions separately.

By applying Eqs.~(\ref{fft-ff}) and (\ref{parFF}) one can see that
there are six independent $F$'s:
\be
F^{++}_{++}\,,\,\,F^{++}_{--}\,,\,\,F^{+-}_{+-}\,,\,\,F^{-+}_{+-}\,,\,
F^{++}_{+-}\,,\,\,F^{+-}_{++}\,.
\ee
These are in principle complex quantities, but $F^{++}_{++}$ and
$F^{++}_{--}$ are clearly moduli squared (see Eq.~\ref{defFF}),
whereas $F^{+-}_{+-}$ and $F^{-+}_{+-}$ are purely imaginary for gluons and
purely real for quarks, as given by Eq.~(\ref{parFF}).
This leaves us with eight independent {\it real} quantities, which are 
directly related to the eight distribution functions defined in 
Eqs.~(\ref{DxY}-\ref{Dunp}) (for quarks) and (\ref{DxY-g}-\ref{Dunp-g}) 
(for gluons), as we are going to show.

\subsection{\label{q-sect} Quark sector}

Let us consider first quark partons. Inserting Eqs.~(\ref{rho-A}) and 
(\ref{fft-ff}) into Eq.~(\ref{defF}), and exploiting the parity relationships 
(\ref{parFF}), yields, for a generic hadronic spin state:   
\bea
\rho^{a/A,S_A}_{++} \hf_{a/A,S_A} &=& \frac{1}{2} \, (1+P_z^a) \,
\hf_{a/A,S_A} = \frac{1}{2} \left( F^{++}_{++} + F^{++}_{--} \right) +
\frac{1}{2} \, P_L^A \left( F^{++}_{++} - F^{++}_{--} \right)
\nonumber \\
&+& P_T^A \, \left[ {\rm Re} F^{++}_{+-} \, \cos(\phi_{S_A} -\phi_a)
+ {\rm Im} F^{++}_{+-} \, \sin(\phi_{S_A} -\phi_a) \right]
\label{rho++}
\eea
\bea
\rho^{a/A,S_A}_{--} \hf_{a/A,S_A} &=& \frac{1}{2} \, (1-P_z^a) \,
\hf _{a/A,S_A} = \frac{1}{2} \left( F^{++}_{++} + F^{++}_{--} \right) -
\frac{1}{2} \, P_L^A \left( F^{++}_{++} - F^{++}_{--} \right)
\nonumber \\
&-& P_T^A \, \left[ {\rm Re} F^{++}_{+-} \, \cos(\phi_{S_A} -\phi_a)  
- {\rm Im} F^{++}_{+-} \, \sin(\phi_{S_A} -\phi_a) \right]
\label{rho--}
\eea
\bea
&& \rho ^{a/A,S_A}_{+-}\hf _{a/A,S_A} = \frac{1}{2} \, (P^a_x-iP_y^a) \,
\hf_{a/A,S_A} = 
i \, {\rm Im} F^{+-}_{++} \,+\, P_L^A \, {\rm Re} F^{+-}_{++} 
\nonumber \\
&& \>\>\>\> 
+ \> \frac{1}{2} \, P_T^A \left[ \left( F^{+-}_{+-} + F^{-+}_{+-} \right) 
\cos(\phi_{S_A} -\phi_a) 
- i \, \left(  F^{+-}_{+-} - F^{-+}_{+-} \right) 
\sin(\phi_{S_A} -\phi_a) \right] \,. \label{rho+-}
\eea

By summing and subtracting Eqs.~(\ref{rho++}) and (\ref{rho--}), one finds
\bea
&&\hf_{a/A,S_A} = \left( F^{++}_{++} + F^{++}_{--} \right) +
2\, P_T^A \, {\rm Im} F^{++}_{+-} \sin(\phi_{S_A} -\phi_a)  
\label{fsa}\\
&&P_z^a\,\hf _{a/A,S_A} = 
P_L^A \left( F^{++}_{++} - F^{++}_{--} \right) +
2\, P_T^A \, {\rm Re} F^{++}_{+-} \cos(\phi_{S_A} -\phi_a) \>,
\label{Pzsa}
\eea
while from the real and imaginary parts of Eq.~(\ref{rho+-}), 
\bea
P_x^a \, \hf_{a/A,S_A} 
&=& 2 \, P^A_L \, {\rm Re} F^{+-}_{++} + P_T^A \, \left( 
F^{+-}_{+-} + F^{-+}_{+-} \right) \, \cos(\phi_{S_A}-\phi_a) 
\label{Pxsa}\\
P_y^a \, \hf_{a/A,S_A} 
&=& -2 \, {\rm Im}  F^{+-}_{++} + P_T^A \, \left( 
F^{+-}_{+-} - F^{-+}_{+-} \right) \, \sin(\phi_{S_A}-\phi_a) \>.
\label{Pysa}
\eea

The two above equations can be written in a compact form (which we shall 
use later) in terms of the parton transverse spin 
\be
P_x^a = P^a_T \cos\phi_{s_a}  \quad\quad\quad
P_y^a = P^a_T \sin\phi_{s_a} \label{qPT},
\ee
where $\phi_{s_a}$ is the azimuthal angle of the polarization vector 
of parton $a$ in its helicity frame. By multiplying Eqs.~(\ref{Pxsa})
and (\ref{Pysa}) respectively by $\cos\phi_{s_a}$ and $\sin\phi_{s_a}$
and summing, one obtains:
\bea 
P^a_T \hf_{a/A,S_A} &=& -2 \, {\rm Im} F^{+-}_{++} \sin\phi_{s_a} +
2 \, P^A_L \, {\rm Re} F^{+-}_{++}  \cos\phi_{s_a} \nonumber \\ 
&\;\;& + \; P_T^A \, \left[ F^{+-}_{+-} \cos(\phi_{s_a} - \phi_{S_A} + \phi_a) 
+ F^{-+}_{+-} \cos(\phi_{s_a} + \phi_{S_A} - \phi_a) \right] \,. \label{PTsa}
\eea
Moreover, one can show that the azimuthal angle of $\bfP^a$ in its 
helicity frame, $\phi_{s_a}$, and the same angle measured in the hadronic 
helicity frame, $\phi^\prime_{s_a}$, are related by
\be
\phi_{s_a}= \phi^\prime_{s_a} - \phi_a + 
{\cal O}\left( \left[ \frac {k_{\perp a}}{x_a \sqrt{s}} \right]^2 \right) \>,
\ee
so that, up to such corrections, Eq.~(\ref{PTsa}) can be written as 
\bea
P^a_T \hf_{a/A,S_A} &=& -2\,{\rm Im} F^{+-}_{++}
                       \sin(\phi^\prime _{s_a} - \phi_a) +
                       2\,P^A_L\,{\rm Re} F^{+-}_{++}
                       \cos(\phi^\prime _{s_a} -\phi_a)
\nonumber \\ &\;\;& + \;
P_T^A\, \left[ F^{+-}_{+-} \cos(\phi^\prime _{s_a}-\phi_{S_A}) +
        F^{-+}_{+-} \cos(\phi^\prime _{s_a}+\phi_{S_A}-2\phi_a) \right] \,.
\label{PTsa-ap}
\eea

Eqs.~(\ref{fsa})-(\ref{Pysa}) express the quark polarizations in term of 
the distribution amplitudes $F$'s and the hadron polarization. One finds 
eight non zero independent soft functions:
\bea
\hf_{a/A} &=& \hf_{a/A,S_L} = \left( F^{++}_{++} + F^{++}_{--} \right) 
\label{A20} \\
\hf_{a/A,S_T} &=& \left( F^{++}_{++} + F^{++}_{--} \right) +
2 \, {\rm Im} F^{++}_{+-} \sin(\phi_{S_A} -\phi_a)  
\label{A21}\\ 
P_x^a \, \hf_{a/A,S_L} &=& 2 \, {\rm Re} F^{+-}_{++} \label{A22}\\ 
P_x^a \, \hf_{a/A,S_T} &=& 
\left( F^{+-}_{+-} + F^{-+}_{+-} \right) \, \cos(\phi_{S_A}-\phi_a) 
\label{A23}\\ 
P_y^a \, \hf_{a/A,S_L} &=& P_y^a \, \hf_{a/A}
= -2 \, {\rm Im}  F^{+-}_{++} \label{A24}\\
P_y^a \, \hf_{a/A,S_T} &=&
-2 \, {\rm Im}  F^{+-}_{++} +
\left( F^{+-}_{+-} - F^{-+}_{+-} \right) \, \sin(\phi_{S_A}-\phi_a)
\label{A25} \\
P_z^a \,\hf _{a/A,S_L} &=& 
\left( F^{++}_{++} - F^{++}_{--} \right) \label{A26}\\
P_z^a \,\hf _{a/A,S_T} &=&
2 \, {\rm Re} F^{++}_{+-} \cos(\phi_{S_A} -\phi_a) \,. \label{A27}
\eea
Notice also that $P_x^a \, \hf_{a/A} = 0$. 

If we fix $\phi_{S_A} = \pi/2$ as done throughout the paper and adopt the 
notations of Eqs.~(\ref{DxY})-(\ref{Dunp}), the above equations read:
\bea
P_x^a \, \hf_{a/A,S_Y}
&=& \Delta \hf_{{s_x}/S_Y} \equiv \Delta \hf_{{s_x}/\ua} =
\left( F^{+-}_{+-} + F^{-+}_{+-} \right) \sin\phi_a \label{A28}\\
P_y^a \, \hf_{a/A,S_Y}
&=& \Delta \hf_{{s_y}/S_Y} \equiv \Delta \hf_{{s_y}/\ua} =
-2\,{\rm Im} F^{+-}_{++} + \left( F^{+-}_{+-} - F^{-+}_{+-} \right)
\cos\phi_a \label{A29}\\
P_z^a \, \hf_{a/A,S_Y}
&=& \Delta \hf_{{s_z}/S_Y} \equiv \Delta \hf_{s_z/\ua} =
2\,{\rm Re} F^{++}_{+-} \sin\phi_a \label{A30}\\
P_x^a \, \hf_{a/A,S_Z}
&=& \Delta \hf_{{s_x}/S_Z} \equiv \Delta \hf_{{s_x}/+} =
2\,{\rm Re} F^{+-}_{++}\label{A31} \\
P_y^a \, \hf_{a/A,S_Z}
&=& \Delta \hf_{{s_y}/S_Z} \equiv \Delta \hf_{{s_y}/+} = \Delta \hf_{{s_y}/A} =
-2\,{\rm Im} F^{+-}_{++} \label{A32}\\
P_z^a \, \hf_{a/A,S_Z}
&=& \Delta \hf_{{s_z}/S_Z} \equiv \Delta \hf_{{s_z}/+} =
\left( F^{++}_{++}  - F^{++}_{--} \right) \label{A33}\\
\hf _{a/A,S_Y} &=& \hf_{a/A} + \frac{1}{2}\,\Delta \hf_{a/S_Y}=
\left( F^{++}_{++}  + F^{++}_{--} \right) + 2\,{\rm Im} F^{++}_{+-}\cos\phi_a  
\,, \label{A34}
\eea
which gives the exact expressions of Eqs.~(\ref{DxY})--(\ref{Dunp}) in 
terms of helicity distribution amplitudes. In particular, Eqs.~(\ref{A32}) 
and (\ref{A34}) allow to obtain the expressions of the Boer-Mulders and 
Sivers functions respectively [see Eqs.~(\ref{Dunp}), (\ref{defsivnoi}) 
and (\ref{parbm})]: 
\bea
\Delta^N \hf_{\aup/A} &=& -2 \, {\rm Im} F^{+-}_{++} \label{b-mda}\\
\Delta^N \hf_{a/\Aup} &=&  4 \, {\rm Im} F^{++}_{+-} \>. \label{sivda}
\eea
Notice also that:
\be
\Delta^- \hf^a_{s_y/S_Y} \equiv \frac 12 \, \left[
\Delta \hf^a_{s_y/\ua} - \Delta \hf^a_{s_y/\da} \right]
= \left( F^{+-}_{+-} - F^{-+}_{+-} \right) \cos\phi_a \>. \label{d-qda}
\ee 

\subsection{\label{gl-sect} Gluon sector}

Thanks to the formal analogy between Eqs.~(\ref{rho-a}) and (\ref{rho-gl})
the expressions of the circular and linear polarizations of the gluons 
in terms of the corresponding helicity distribution amplitudes are 
closely analogous to those obtained for quarks in the previous subsection. 
One should only pay attention to the parity properties appropriate for 
spin 1 gluons and remember that the $F$'s are now the helicity distribution 
amplitudes for the $A \to g + X$ process.   

One finds that Eqs.~(\ref{rho++}) and (\ref{rho--}) hold true also for 
gluons, while Eq.~(\ref{rho+-}), due to the different parity relationships, 
changes into:
\bea
&& \rho ^{g/A,S_A}_{+-}\hf _{g/A,S_A} =
\frac{1}{2}(\IoneG-i\ItwoG)\hf _{g/A,S_A} =
{\rm Re} F^{+-}_{++} \,+\, i \, P_L^A \, {\rm Im} F^{+-}_{++}
\nonumber \\
&& \>\>\>
- \, \frac{i}{2} \, P_T^A \left[ (F^{+-}_{+-} + F^{-+}_{+-}) \sin(\phi_{S_A} 
-\phi_a) + \, i \, (F^{+-}_{+-} - F^{-+}_{+-}) 
\cos(\phi_{S_A} -\phi_a) \right] \,, \label{rho+-g} 
\eea 
where $F^{+-}_{+-}$ and $F^{-+}_{+-}$ are now purely imaginary quantities. 

As a consequence, Eqs.~(\ref{fsa}) and (\ref{Pzsa}) keep describing the 
distributions of unpolarized or longitudinally polarized gluons inside a 
polarized hadron, while Eqs.~(\ref{Pxsa}) and (\ref{Pysa}) modify into:
\bea
\IoneG \fgAS &=& 2 \, {\rm Re} F_{++}^{+-} + P_T^A \, 
{\rm Im} \left(F_{+-}^{+-} + F_{+-}^{-+} \right) \,
\sin(\phi_{S_A}-\phi_a) \label{rerho+-g} \\
\ItwoG \fgAS &=&-2 \, P_L^A \,{\rm Im} F_{++}^{+-}
- P_T^A \, {\rm Im} \left(F_{+-}^{+-} - F_{+-}^{-+} \right) \,
\cos(\phi_{S_A}-\phi_a) \,.\label{imrho+-g}
\eea

Eqs.~(\ref{A20})--(\ref{A27}) now become:  
\bea
\hf_{g/A} &=& \hf_{g/A,S_L} = \left( F^{++}_{++} + F^{++}_{--} \right) 
\label{A41} \\
\hf_{g/A,S_T} &=& \left( F^{++}_{++} + F^{++}_{--} \right) +
2 \, {\rm Im} F^{++}_{+-} \sin(\phi_{S_A} -\phi_a)  
\label{A42} \\ 
\IoneG \, \hf_{g/A,S_L} &=& \IoneG \, \hf_{g/A} = 
2 \, {\rm Re} F^{+-}_{++} \label{A43} \\ 
\IoneG \, \hf_{g/A,S_T} &=& 2 \, {\rm Re} F^{+-}_{++} + {\rm Im}
\left( F^{+-}_{+-} + F^{-+}_{+-} \right) \, \sin(\phi_{S_A}-\phi_a) 
\label{A44} \\ 
\ItwoG \, \hf_{g/A,S_L} &=& -2 \, {\rm Im}  F^{+-}_{++} \label{A45} \\
\ItwoG \, \hf_{g/A,S_T} &=&
- \, {\rm Im} \left( F^{+-}_{+-} - F^{-+}_{+-} \right) \, 
\cos(\phi_{S_A}-\phi_a) \label{A46} \\
P_z^g \,\hf _{g/A,S_L} &=& 
\left( F^{++}_{++} - F^{++}_{--} \right) \label{A47} \\
P_z^g \,\hf _{g/A,S_T} &=&
2 \, {\rm Re} F^{++}_{+-} \cos(\phi_{S_A} -\phi_a) \>, 
\label{A48}
\eea
while, choosing $\phi_{S_A} = \pi/2$ and following the notation of 
Eqs.~(\ref{DxY-g})--(\ref{Dunp-g}), we have:
\bea
\IoneG \, \hf_{g/A,S_Y} &=& \Delta \hf_{\Ione/S_Y}^g = 
\Delta \hf_{{\Ione}/\ua}^g = 2\,{\rm Re} F^{+-}_{++} +
{\rm Im} \left( F^{+-}_{+-} + F^{-+}_{+-} \right) \cos\phi_a\label{A49} \\
\ItwoG \, \hf_{g/A,S_Y} &=& \Delta \hf_{\Itwo/S_Y}^g = 
\Delta \hf_{{\Itwo}/\ua}^g =
- {\rm Im} \left( F^{+-}_{+-} - F^{-+}_{+-} \right) \sin\phi_a \label{A50} \\
P_z^g \, \hf_{g/A,S_Y} &=& \Delta \hf_{s_z/S_Y}^g = \Delta \hf_{{s_z}/\ua}^g =
2\,{\rm Re} F^{++}_{+-} \sin\phi_a \label{A51} \\
\IoneG \, \hf_{g/A,S_Z} &=& \Delta \hf_{\Ione/S_Z}^g = 
\Delta \hf_{\Ione/+}^g = \Delta \hf_{\Ione/A
}^g =  2\,{\rm Re} F^{+-}_{++} \label{A52} \\
\ItwoG \, \hf_{g/A,S_Z} &=& 
\Delta \hf_{\Itwo/S_Z}^g = \Delta \hf_{\Itwo/+}^g = -2\,{\rm Im} F^{+-}_{++}
\label{A53} \\
P_z^g \,\hf _{g/A,S_Z} &=& \Delta \hf_{s_z/S_Z}^g = \Delta \hf_{{s_z}/+}^g =
\left( F^{++}_{++} - F^{++}_{--} \right) \label{A54} \\
\hf _{g/A,S_Y} &=& \hf_{g/A} + \frac{1}{2}\,\Delta \hf_{g/\ua}=
\left( F^{++}_{++} + F^{++}_{--} \right)  + 2\, {\rm Im}F^{++}_{+-} 
\cos\phi_a \,. \label{A55}
\eea

The Sivers function (\ref{sivda}) can exist also for gluons, while the 
``Boer-Mulders-like'' function is given by
\be
\IoneG \hf_{g/A} = \Delta \hf_{\Ione/A}^g = \Delta \hf_{\Ione/+}^g
= 2 \, {\rm Re} F^{+-}_{++} \>. \label{A57}
\ee 
Finally, in analogy to Eq.~(\ref{d-qda}):
\be
\Delta^- \hf^g_{\Ione/S_Y} \equiv \frac 12 \, \left[
\Delta \hf^g_{\Ione/\ua} - \Delta \hf^g_{\Ione/\da} \right]
= {\rm Im} \left( F^{+-}_{+-} + F^{-+}_{+-} \right) \cos\phi_a \>. 
\label{d-gda}
\ee 

\section{\label{Mulders-quark} Relations between different notations}

\subsection{Quark distribution functions}

Let us compare our notations with those used in the formalism of the 
Amsterdam group~\cite{amst} (see also Ref.~\cite{brd}), which is widely 
used. In this formalism the main object, corresponding to our 
$\hat F_{\la^{\,}_A,\la^{\prime}_A}^{\la^{\,}_a,\la^{\prime}_a}(x_a, 
\bfk_{\perp a})$, is the $\Phi(x_a,\bfk_{\perp a})$ correlator
\bea
\Phi(x_a,\bfk_{\perp a}) &=& \frac{1}{2} \left[
 f_1\slash{n}_+ +
f_{1T}^\perp \frac{\epsilon _{\mu\nu\rho\sigma}
\gamma^\mu n_+^\nu k_{\perp a}^\rho
(P_T^A)^\sigma}{M} +
\left( P_L^Ag_{1L} +
\frac{\bfk_{\perp a}\cdot\bfP_T^A}{M} g_{1T}^\perp \right)
\gamma ^5 \slash{n}_+
\right. \nonumber \\ &+&
\left. h_{1T} \,i\sigma_{\mu\nu} \gamma^5 n_+^\mu (P_T^A)^\nu +
\left( P_L^A h_{1L}^\perp + \frac{\bfk_{\perp a}
\cdot\bfP_T^A}{M}h_{1T}^\perp\right)
\frac{i\sigma_{\mu\nu} \gamma^5 n_+^\mu k_{\perp a}^\nu}{M} \right.
\nonumber \\
&+&\left. h_1^\perp \frac{\sigma _{\mu\nu} k_{\perp a}^\mu n_+^\nu}{M}
\right]\,.
\label{Phi}
\eea

By appropriate Dirac projections $\Phi^{[\Gamma]} = {\rm Tr}(\Gamma \Phi)$ 
one can single out the various sectors of distribution functions.
In particular, $\Gamma=(n_-)_\alpha \gamma^\alpha/2$ projects out the
$f_1$ sector ({\it i.e.} all the distribution functions relative to an 
unpolarized quark), namely the usual unpolarized distribution function
$f_1^a(x_a,k_{\perp a})$ and the Sivers function 
$f_{1T}^{\perp a}(x_a,k_{\perp a})$:
\be
{\rm Tr}\left( \frac{\slash{n}_-}{2} \Phi \right) =
f_1-P_T^A \frac{k_{\perp a}}{M}\sin(\phi_{S_A}-\phi_a)\,f_{1T}^\perp\,,
\label{f-sect}
\ee
where $M$ is the proton mass and $n_\pm = 1/\sqrt{2} \,(1,0,0,\pm 1)$.

Similarly, the projection operator
$\Gamma=(n_-)_\alpha \gamma^\alpha \gamma^5/2$ gives the $g_1$ sector
({\it i.e.} the distribution functions corresponding to a longitudinally 
polarized quark), namely the helicity distribution function 
$g_{1L}^a(x_a,k_{\perp a})$ and the number density of longitudinally 
polarized partons $a$ in a transversely polarized hadron $A$, called 
$g_{1T}^{\perp a}(x_a,k_{\perp a})$:
\be
{\rm Tr}\left( \frac{\slash{n}_-}{2}\,\gamma^5\, \Phi \right) =
P_L^A \,g_{1L} + P_T^A \frac{k_{\perp a}}{M}
\cos(\phi_{S_A} -\phi_a)\,g_{1T}^\perp\,.
\label{g-sect}
\ee

Finally, to obtain the $h_1$ sector ({\it i.e.} the distribution functions 
relative to a transversely polarized quark), we have to apply the projector
$\Gamma=\frac{1}{2} i \sigma_{\mu\nu}(n_-)^\mu \frac{(P_T^A)^\nu}{2} \gamma^5$:
\bea
{\rm Tr}\left( \frac{1}{2} i \sigma_{\mu\nu}(n_-)^\mu \frac{(P_T^A)^\nu}{2}
\gamma^5\,
\Phi \right) &=&
\,P_T^A \left[\cos(\phi_{S_A}-\phi^\prime _{s_a}) \, h_{1} +
\frac{k_{\perp a}^2}{2M^2}\cos(2\phi_a-\phi_{S_A}-\phi^\prime _{s_a})\,
h_{1T}^\perp \right] 
\nonumber \\
&+&\frac{k_{\perp a}}{M}\cos(\phi_a-\phi^\prime _{s_a}) \, P_L^A \, 
h_{1L}^\perp - \frac{k_{\perp a}}{M} \sin(\phi^\prime _{s_a}-\phi_a) 
h_1^\perp \>, \label{h-sect}
\eea
with 
\be
h_1 = h_{1T} + \frac{k_{\perp a}^2}{2 M^2} \, h_{1T}^{\perp} \>. \label{h1h1T}
\ee

The relations between the
$F^{\la_a^{\,}, \la_a^\prime}_{\la_A^{\,}, \la_A^\prime}$ inclusive cross
sections and the Amsterdam group distribution functions can
straightforwardly be derived by comparing Eqs.~(\ref{f-sect}), 
(\ref{g-sect}) and (\ref{h-sect}) with Eqs.~(\ref{fsa}), (\ref{Pzsa}) 
and (\ref{PTsa-ap}) respectively, obtaining:
\bea
f_1(x_a,k_{\perp a}) &=& F^{++}_{++} + F^{++}_{--} = \hf_{a/A} \label{f1} \\
\frac{k_{\perp a}}{M} \> f_{1T}^\perp(x_a,k_{\perp a}) 
&=& -2\,{\rm Im} F^{++}_{+-} \label{f1Tp} \\
g_{1L}(x_a,k_{\perp a}) &=& F^{++}_{++} - F^{++}_{--} \label{g1} \\
\frac{k_{\perp a}}{M} \> g_{1T}^\perp(x_a,k_{\perp a}) 
&=& 2\,{\rm Re} F^{++}_{+-} \label{g1Tp} \\
\frac{k_{\perp a}}{M} \> h_{1L}^\perp(x_a,k_{\perp a}) 
&=& 2\,{\rm Re} F^{+-}_{++} \label{h1Lp} \\
\frac{k_{\perp a}}{M} \> h_{1}^\perp(x_a,k_{\perp a}) 
&=&  2\,{\rm Im} F^{+-}_{++} \label{h1p} \\
h_1(x_a,k_{\perp a}) &=& F^{+-}_{+-} \label{h1T} \\
\left( \frac{k_{\perp a}}{M} \right)^2 \> h_{1T}^\perp(x_a,k_{\perp a}) 
&=& 2\,F^{-+}_{+-} \label{h1Tp} \>.
\eea

Notice that, according to the most general forward behaviour of helicity
amplitudes (see, {\it e.g.}, Eq.~(4.3.1) on page 79 of Ref.~\cite{elliot}),
one should have the minimal requirement: 
\be
F^{\la_a^{\,}, \la_a^\prime}_{\la_A^{\,}, \la_A^\prime}(x_a, k_{\perp a} =0) 
\sim (k_{\perp a})^{|\la_A - \la_a + \la^\prime_A -\la^\prime_a|} \>,
\label{ktdep}
\ee
which is explicit in the above equations. The proton mass $M$ is assumed in 
Eq.~(\ref{Phi}) as a reasonable scale for the intrinsic motion $k_\perp$.

Combining Eqs.~(\ref{f1})--(\ref{h1Tp}) with Eqs.~(\ref{A20})--(\ref{A27}) 
one can obtain the relationships between the Amsterdam functions and the 
quark polarizations. Using Eqs. (\ref{f1Tp}), (\ref{h1Lp}), (\ref{h1p}), 
(\ref{g1}) and (\ref{g1Tp}) respectively into Eqs.~(\ref{A21}), (\ref{A22}), 
(\ref{A24}), (\ref{A26}) and (\ref{A27}), yields:
\bea
\hf_{a/A,S_T} - \hf_{a/A,-S_T} &=& \Delta \hf_{a/S_T}(x_a,\bfk_{\perp a}) =
-2\,\frac{k_{\perp a}}{M} \sin(\phi_{S_A}-\phi_a) \,
f_{1T}^\perp(x_a,k_{\perp a}) \\
P^a_x \, \hf_{a/A,S_L} &=& \Delta \hf_{s_x/+}(x_a,\bfk_{\perp a}) =
\frac{k_{\perp a}}{M} \, h_{1L}^\perp(x_a,k_{\perp a}) \\
P^a_y \, \hf_{a/A,S_L} = P^a_y \, \hf_{a/A} &=& 
\Delta \hf_{s_y/A}(x_a,\bfk_{\perp a}) =
-\frac{k_{\perp a}}{M} \, h_{1}^\perp(x_a,k_{\perp a}) \\
P^a_z \, \hf_{a/A,S_L} &=& \Delta \hf_{s_z/+}(x_a,\bfk_{\perp a}) 
= g_{1L}(x_a,k_{\perp a}) \\
P^a_z \, \hf_{a/A,S_T} &=& \Delta \hf_{s_z/S_T}(x_a,\bfk_{\perp a}) =
\frac{k_{\perp a}}{M}\cos(\phi_{S_A}-\phi_a)\,g_{1T}^\perp(x_a,k_{\perp a}) \,,
\eea
which shows that the functions $f_{1T}^\perp,  h_{1L}^\perp, h_{1}^\perp,
g_{1L}$ and $g_{1T}^\perp$ have a direct physical interpretation in terms of 
corresponding polarized quark distributions. 

Instead, insertion of Eqs.~(\ref{h1p})--(\ref{h1Tp}) and (\ref{h1h1T})
into Eqs.~(\ref{A23}) and (\ref{A25}) gives 
\bea
P_x^a \, \hf_{a/A,S_T} &=& \Delta \hf_{s_x/S_T}(x_a,\bfk_{\perp a}) 
\nonumber \\ 
&=& \left[ h_{1T}(x_a,k_{\perp a}) + \frac{k_{\perp a}^2}{M^2} \, 
h_{1T}^\perp(x_a,k_{\perp a}) \right] \, \cos(\phi_{S_A}-\phi_a) \label{B20}\\ 
P_y^a \, \hf_{a/A,S_T} &=& \Delta \hf_{s_y/S_T}(x_a,\bfk_{\perp a}) 
\nonumber \\
&=& - \frac{k_{\perp a}}{M} \, h_1^\perp (x_a,k_{\perp a}) +
h_{1T}(x_a,k_{\perp a}) \,\sin(\phi_{S_A}-\phi_a) \,, \label{B21}
\eea 
which shows that $h_{1T}$ and $h_{1T}^\perp$ are combinations of quark 
polarized distributions. 

\subsection{\label{Mulders-gluons} Gluon distribution functions}

In Ref.~\cite{mr01} Mulders and Rodriguez discussed the twist-two transverse 
momentum dependent gluon distribution functions for spin-1/2 hadrons. 
Their notation is different from ours, and it is worth mentioning the 
relations which link the two different formalisms.

Naming conventions in Ref.~\cite{mr01} are set as follows:
$G$ and $\Delta G$ indicate gluon distribution functions which are diagonal 
in the gluon helicities, {\it i.e.} correspond to either unpolarized ($G$) or 
circularly polarized ($\Delta G$) gluons. $H$ and $\Delta H$ indicate gluon 
distribution functions which correspond to linearly polarized gluons in either
unpolarized or polarized hadrons respectively. As for the quark distribution 
functions, a $T$ or $L$ subscript indicates that the parent hadron is either 
transversely or longitudinally polarized, and a $\perp$ superscript shows an 
explicit dependence of the distribution function on the gluon intrinsic 
transverse momentum.

Indeed, eight such functions exist:
\begin{itemize}
\item
$G$ is the usual distribution function of
unpolarized gluons inside unpolarized hadrons, corresponding to
$\hf_{g/A} = F^{++}_{++} + F^{++}_{--}$, Eq.~(\ref{A41});
\item
$\Delta G_L$ is the distribution function of circularly polarized gluons 
inside a longitudinally polarized hadron $A$, corresponding to
$\Delta \hf^g_{s_z/+} = F^{++}_{++} - F^{++}_{--}$, Eqs.~(\ref{A47})
and (\ref{A54});
\item
$G_T$ is the distribution function of unpolarized gluons inside a transversely 
polarized hadron, {\it i.e.} the gluon Sivers function, corresponding to 
$\Delta^N \hf_{g/\Aup} = 4\,{\rm Im} F^{++}_{+-}$, Eq.~(\ref{sivda});
\item
$\Delta G_T$ is the distribution function of circularly polarized gluons
inside a transversely polarized hadron, corresponding to
$\Delta \hf^g_{s_z/S_T} = 2\,{\rm Re} F^{++}_{+-}$, Eqs.~(\ref{A48})
and (\ref{A51});
\item
$H^\perp$ is the distribution function of linearly polarized gluons in
unpolarized hadrons, which corresponds to
$\Delta \hf^g_{\Ione/A} = {\rm Re} F^{+-}_{++}$, Eqs.~(\ref{A43}) and 
(\ref{A52});
\item
$H_L^\perp$ is the distribution function of linearly polarized gluons in
longitudinally polarized hadrons, which corresponds to
$\Delta \hf^g_{\Itwo/+} = {\rm Im} F^{+-}_{++}$, Eqs.~(\ref{A45}) 
and (\ref{A53});
\item
$\Delta H_T$ and  $\Delta H_T ^\perp$ are related to the distribution
function of linearly polarized gluons in transversely  polarized hadrons,
$\Delta H_T^\prime = \Delta H_T - (k_{\perp g}^2/2M^2) \Delta H_T ^\perp$.
In this case, it is difficult to find a precise relation between the two
formalisms, but we can say that $\Delta H_T$ and  $\Delta H_T ^\perp$
play the same role as $F^{+-}_{+-}$ and $F^{+-}_{-+}$, similarly to the quark
case [see Eqs.~(\ref{A28}), (\ref{A29}), (\ref{B20}) and (\ref{B21})].
\end{itemize}
Notice that Eq.~(\ref{ktdep}) is valid for gluons as well as for quarks.

\section{Helicity frames}

Our physical observables are computed in the $AB$ c.m. frame (overall hadronic 
frame) with axes denoted by $X_{cm}, \, Y_{cm}, \, Z_{cm}$. The helicity 
frame of a particle with momentum $\bfp$ along the direction $\hat{\bfp} = 
(\sin\theta\cos\varphi, \, \sin\theta\sin\varphi, \, \cos\theta)$ -- {\it as 
defined in the hadronic frame} -- can be reached by performing on the overall 
frame the rotations \cite{elliot} 
\be 
R(\varphi, \theta, 0) = R_{Y'}(\theta) \, R_{Z_{cm}}(\varphi) \>.
\ee
The first is a rotation by an angle $\varphi$ around the $Z_{cm}$-axis and 
the second is a rotation by an angle $\theta$ around the new (that is, 
obtained after the first rotation) $Y'$-axis. 

This results in the helicity frames with axes along the following directions 
(expressed in the hadronic frame):
\be 
\hat{\bfX}_A = \hat{\bfX}_{cm} \quad\quad\quad \hat{\bfY}_A = 
\hat{\bfY}_{cm} \quad\quad\quad \hat{\bfZ}_A = \hat{\bfZ}_{cm} 
\ee
for a hadron A moving along +$\hat{\bfZ}_{cm}$,  
\be
\hat{\bfX}_B = \hat{\bfX}_{cm} \quad\quad\quad \hat{\bfY}_B = 
- \hat{\bfY}_{cm} \quad\quad\quad \hat{\bfZ}_B = - \hat{\bfZ}_{cm} 
\label{helB}
\ee
for a hadron B moving along $-\hat{\bfZ}_{cm}$,
\be 
\hat{\bfx} = \hat{\bfy} \times \hat{\bfz} \quad\quad\quad
\hat{\bfy} = \frac{\hat{\bfZ}_{cm} \times \hat{\bfp}}
{|\hat{\bfZ}_{cm} \times \hat{\bfp}|} = 
\hat{\bfZ}_{cm} \times \hat{\bfk}_\perp 
\quad\quad\quad \hat{\bfz} = \hat{\bfp} \label{help}
\ee
for a generic particle $\bfp$. Notice that $\hat{\bfk}_\perp$ is the unit 
transverse component -- with respect to the $Z_{cm}$-direction -- of $\bfp$, 
and that it lies in the $(xz)$ plane.

\end{document}